\documentclass[12pt]{article}
\usepackage{amssymb}
\usepackage{amsfonts}
\usepackage{amsmath}

\newtheorem{theorem}{Theorem}

\newtheorem{claim}{Claim}

\newtheorem{corollary}{Corollary}

\newtheorem{lemma}{Lemma}

\newtheorem{proposition}{Proposition}
\newtheorem{remark}{Remark}

\begin{document}

\title{School Choice with Multiple Priorities\thanks{%
This work was supported by JSPS KAKENHI Grant Numbers 20K01675, 22K01402 and
24K04932.}}
\author{Minoru Kitahara \and Yasunori Okumura\thanks{%
Corresponding author. Department of Logistics and Information Engineering,
TUMSAT, 2-1-6, Etchujima, Koto-ku, Tokyo, 135-8533 Japan.
Phone:+81-3-5245-7300. Fax:+81-3-5245-7300. E-mail:
okuyasu@gs.econ.keio.ac.jp}}
\maketitle

\begin{center}
\textbf{Abstract}
\end{center}

This study considers a model where schools may have multiple priority orders
on students, which may be inconsistent with each other. For example, in
school choice systems, since the sibling priority and the walk zone priority
coexist, the priority orders based on them would be conflicting. We
introduce a weaker fairness notion called M-fairness to examine such
markets. Further, we focus on a more specific situation where all schools
have only two priority orders, and for a certain group of students, a
priority order of each school is an improvement of the other priority order
of the school. An illustrative example is the school choice matching market
with a priority-based affirmative action policy. We introduce a mechanism
that utilizes the efficiency adjusted deferred acceptance algorithm and show
that the mechanism satisfies properties called responsiveness to
improvements and improved-group optimally M-stability, which is stronger
than student optimally M-stability.


\newpage

\section{Introduction}

In this study, we consider a matching model where schools are allowed to
have multiple priority orders on students, which may be inconsistent with
each other. An illustrative example of such a situation is evident in the
old school choice system in Boston city, where both sibling priority and
walk zone priority coexist (Abdulkadiro\u{g}lu et al. (2005, 2006)).
Conflicts may arise between the priority order based on sibling status and
that based on walk zone status, particularly when one student has sibling
priority only, and another has walk zone priority only. Such a conflict does
not ensure the existence of a nonwasteful matching that respects both
priority orders simultaneously. Thus, in such school choice systems, by
determining which priority is more important, one single linear priority
order for each school is employed to determine the assignments. On the other
hand, this study proposes a mechanism that leads to a more efficient outcome
compared to a standard one using a profile of linear priority orders.

Moreover, recently, the walk zone priority was suppressed in the school
choice system in Boston city (Dur et al. (2018)), but, as discussed by Dur
et al. (2019, Section 5), just suppressing it weakens the fairness aspect of
the result with not improving or sometimes worsening it in terms of
efficiency. Our method improves matching efficiency by relaxing the two
priorities while still considering them, rather than completely suppressing
one of them.

We consider a general situation where each school is allowed to have
multiple priority orders such as one is based on sibling priority and
another is on walk zone priority. In that case, there may be no non-wasteful
matching that respects to all priority orders. Thus, we introduce a weaker
fairness notion (than the usual one) called M-fairness, wherein the
violation of a priority order of a school between two students is allowed if
the violation is justified by another priority of the school between these
students.

To clarify the distinction, we refer to the priority order profile in the
typical case where each school has only one priority as a single priority
order profile,\ and in the case where at least one school has two or more
priority orders, we refer to the profile as a multiple priority orders
profile.\ We construct the single priority order profile from a multiple
priority orders profile such that a matching is M-fair for the multiple
priority order profile if and only if it is fair (in a usual sense) for the
single priority order profile.

Moreover, we introduce the following computationally efficient mechanism.
First, a single priority order profile is constructed from a multiple
priority orders profile. Second, we use a variant of the efficiency adjusted
deferred acceptance algorithm (hereafter EADA algorithm), which is
introduced by Kesten (2010), with the constructed profile and a single
priority order profile for which the status quo matching is stable. This
mechanism results in a student optimal M-fair (stable) matching for the
multiple priority orders profile that weakly Pareto dominates any stable
matching for a single priority order profile, such as one based only on
sibling priority.

Next, we consider the group optimality of M-stable matchings. We say that a
matching is Pareto dominated \textit{for a group} by another matching if no
student in the group prefers the latter to the former and some in the group
prefers the former to the latter. A group optimal M-stable matching for a
multiple priority orders profile is an M-stable matching for that profile if
it is not Pareto dominated for a certain group by any M-stable matchings for
it. We focus on a specific multiple priority orders profile, where all
schools have two priority orders, and for a certain group of students, a
priority order of each school is an improvement of the other priority order.
We show that our mechanism is improved-group optimally (also whole student
optimally) M-stable for the multiple priority orders profile; that is, the
mechanism results in an M-stable matching that is not Pareto dominated for
the improved-group by any M-stable matchings.

As an illustrative example, we consider the market where the students are
divided into two groups: majority students and (under-represented) minority
students. In such markets, authorities sometimes aim to\ provide minority
students with greater educational opportunities through affirmative action
policies. Among these policies, our focus is on the priority-based
affirmative action policy, because in this context, there are two priority
orders for each school that should be considered.\footnote{%
For example, in Chinese college admissions, the minority students are
awarded bonus points in the national college entrance examination. See, for
example, Wang (2009) on the policy in detail.} First, each school has an
original priority order determined regardless of whether a student is
majority or minority student; e.g., it is determined only by the test score.
Second, each school also has an adjusted priority order aimed at improving
educational opportunities for minority students. In this case, the adjusted
priority order of each school would be an improvement of the original one
for minority students.

However, Kojima (2012) shows that any stable matching for an improvement
profile may be Pareto dominated for the minority students by a stable
matching for the original one. On the other hand, our mechanism leads a
student optimal M-stable matching for the profile of the sets of the two
priority orders of schools that weakly Pareto dominates any stable matching
for the improvement profile. Thus, although the result of the mechanism may
violate some (improved) priorities of minority students, in that case, some
minority students are better off compared to any stable matching for the
improvement profile. Moreover, the mechanism achieves the minority
optimality and thus the result is never Pareto dominated for the minority
students by any stable matching for the original profile.

We also consider two improvements of an original priority profile such that
one profile more improves the original profile than the other. We say that
the mechanism is responsive to improvements if its result with the more
improved profile (and the original one) is not Pareto dominated for the
improved group by that with the less improved profile (and the original
one). We show that our mechanism is responsive to improvements.

\section{Related literature}

In the existing literature on matching theory, many researchers, including
Gale and Shapley (1962), Roth (1982), Balinski and S\"{o}nmez (1999),
Abdulkadiro\u{g}lu and S\"{o}nmez (2003), have emphasized the trade-off
between efficiency and stability in matchings. Consequently, many previous
studies have focused on methods to improve the efficiency of matching
outcomes while considering fairness to some extent.

First, in the studies including Abdulkadiro\u{g}lu et al. (2020), Do\u{g}an
and Ehlers (2021), Tang and Zhang (2021), and Kwon and Shorrer (2023), the
focus is not only on whether justified envy exists, but rather on the extent
to which it exists. They mainly consider the efficient matchings that is
justified envy minimal (according to several definitions).

Second, there are many studies that consider methods to improve efficiency
by allowing the violation of specific priorities. In studies including
Kesten (2010), Morrill (2015), Alcalde and Romero-Medina (2017), Alva and
Manjunath (2019), Ehlers and Morrill (2020), Troyan et al. (2020), and Reny
(2022), whether a specific priority violation is allowed in a matching
depends endogenously on the assignment within that matching. On the other
hand, in studies by Dur et al. (2019), Fang and Yasuda (2021), and Kitahara
and Okumura (2021, 2024), the set of priorities that are allowed to be
violated is exogenously given. In our approach, the set of priorities for
each school that are allowed to be violated is determined by the
(exogenously given) multiple priority orders of the school.

Many of the studies introduced above utilize the EADA algorithm originally
introduced by Kesten (2010). See, for example, Cerrone et al. (2024), for a
survey on that algorithm. In this study, we also employ a variant of EADA
algorithm \`{a} la Tang and Yu (2014), which is a modification of the
original one. Tang and Yu (2014) mention that the EADA algorithm results in
a student optimal stable matching (hereafter SOSM) for a profile of weak
priority orders. In this study, from necessity, we show that the algorithm
also results in an SOSM for a profile of priority orders that may not be
weak orders but partial orders.

Kitahara and Okumura (2021) show that an algorithm of the stable improvement
cycle class, which is a modification of the algorithms introduced by Ergil
and Ergin (2008) and Dur et al. (2019), also results in an SOSM for a
profile of partial order binary relations. Therefore, we can derive a
student optimal M-stable matching by using an algorithm in the stable
improvement cycle class. However, our mechanism utilizing the EADA algorithm
has other remarkable characteristics.

Some previous studies, such as Che et al. (2019a, b), Dur et al. (2019),
Kitahara and Okumura (2020, 2021, 2022, 2024), and Kuvalekar (2023), discuss
matching models where the preference or priority orders of some agents are
not weak orders but partial orders. In our analysis, we construct a partial
order from a set of weak priority orders of each school and derive a
student-optimal M-stable matching by using the profile of the partial orders
and the EADA algorithm.

We apply our approach to the school choice problem with affirmative actions,
which is discussed by many previous studies. See, for example, Abdulkadiro%
\u{g}lu and Grigryan (2024) for a comprehensive survey. Among them, we
consider a priority-based affirmative action policy discussed by Kojima
(2012), Afacan and Salman (2016), Jiao and Tian (2019), Jiao and Shen
(2021), Jiao et al. (2022) and Dur and Xie (2023).\footnote{%
For other types of affirmative action, there are quota-based affirmative
action and reserve-based affirmative action. Moreover, Kitahara and Okumura
(2020, 2023) propose a hybrid approach that combines these three types of
affirmative action.}

Kojima (2012) firstly discusses the responsiveness for the priority-based
affirmative action of the mechanisms and shows that it is incompatible with
stability.\footnote{%
The responsiveness to improvements is related to the respect for
improvements property, as introduced by Balinsky and S\"{o}nmez (1999).
However, the latter property considers the improvement for one student only.
Hirata et al. (2022) consider the respect for group improvements property,
which is more related to ours. They show that any stable mechanism fails to
satisfy the property.} Hafalir et al. (2013) and Do\u{g}an (2016) discuss it
further but focus only on other types of affirmative action: the quota-based
and the reserve-based affirmative action policies.

Jiao and Shen (2021) show that although the EADA mechanism with full
consenting is not responsive to the priority-based affirmative action, that
with only minorities consenting is responsive to it. However, the resulting
matching of the latter mechanism violates some priorities of minority
students. We also utilize an EADA algorithm and show that our mechanism is
stable in our sense as well as responsive to the priority-based affirmative
action. Jiao et al. (2022) derive a condition under which the deferred
acceptance mechanism is responsive to the priority-based affirmative action.
However, as they mentioned, this condition is highly restrictive.

Dur and Xie (2023) weaken the notion of responsiveness to escape from the
impossibility result. They also introduce a responsiveness notion that is
stronger than the usual one and derive a sufficient condition that the
student-proposing deferred acceptance mechanism satisfies it. Our
responsiveness notion is also stronger than usual one discussed by the
previous studies, but we consider no restriction on priority orders and
weaker stability notion.

Finally, there are several recent studies that consider a matching model
where agents are allowed to have multiple preference orders. See, for
example, Chen et al. (2018), Miyazaki and Okamoto (2019) and Aziz et al.
(2020). They consider a one-to-one matching problem, but each agent in
either side may have multiple preference or priority orders. On the other
hand, we consider a general one-to-many matching problem, but the agents in
only one side can have multiple priority orders. Moreover, they consider a
stronger stability notion; e.g., a matching is stable if it is stable in a
usual sense for all preference profiles. On the other hand, we consider a
weaker stability notion than theirs.

\section{Model}

First, we define the following properties of binary relations. Let $B$ be a
binary relation on a set $X$ is

\begin{description}
\item \textbf{asymmetric} if $\left( x,y\right) \in B$ implies $\left(
y,x\right) \notin B$, for all $x,y\in X,$

\item \textbf{total}\textit{\ }if\textit{\ }$x\neq y$ implies\textit{\ }$%
\left( x,y\right) \in B$ or $\left( y,x\right) \in B$ for all $x,y\in X$,

\item \textbf{transitive} if $\left( x,y\right) \in B$ and $\left(
y,z\right) \in B$ imply $\left( x,z\right) \in B$, for all $x,y,z\in X,$

\item \textbf{negatively} \textbf{transitive}\textit{\ }if $\left(
x,y\right) \notin B$ and $\left( y,z\right) \notin B$ imply $\left(
x,z\right) \notin B$, for all $x,y,z\in X$,

\item \textbf{acyclic }if for all $K\in \left\{ 1,2,\cdots \right\} $ and
for all $x_{0},x_{1},\cdots ,x_{K}\in X$, $\left( x_{k-1},x_{k}\right) \in B$
and $\left( x_{k},x_{k-1}\right) \notin B$ for all $k\in \left\{ 1,\cdots
,K\right\} $ imply $\left( x_{K},x_{0}\right) \notin B$.
\end{description}

See Duggan (1999) on the properties in detail. Moreover, if $B$ is
transitive and asymmetric, then it is acyclic.

In this study, our primary focus is on asymmetric binary relations. A binary
relation on $X$ is called a (strict) \textbf{partial order} on $X$ if it is
asymmetric and transitive. A partial order on $X$ is called a (strict) 
\textbf{weak order} on $X$ if it is negatively transitive. A weak order on $%
X $ is called a (strict) \textbf{linear order} on $X$ if it is total.
Moreover, we let $\Pi (B)$ represent an asymmetric part of a binary relation 
$B$; that is, $\left( x,y\right) \in \Pi (B)$ if and only if $\left(
x,y\right) \in B$ and $\left( y,x\right) \notin B$.

If $B$ is a partial order but not a weak order, then $\left( x,y\right)
\notin B$ and $\left( y,z\right) \notin B$, but $\left( x,z\right) \in B$.
Thus, even if two pairs of elements $x,y$ and $y,z$ are incomparable (or
tied), $x$ and $z$ may be comparable (or not tied).

If $B$ is a linear order on $X$, then we simply write 
\begin{equation*}
B:\text{ }x_{0}\text{ }x_{1}\text{ }\ldots \text{ }x_{\left\vert
X\right\vert -1},
\end{equation*}%
representing that $\left( x_{k-1},x_{k}\right) \in B$ for all $k=1,\cdots
,\left\vert X\right\vert -1$. On the other hand, if $B^{\prime }$ is a weak
order on $X$ and $\left( x_{1},x_{2}\right) ,\left( x_{2},x_{1}\right)
\notin B^{\prime }$, then we simply write 
\begin{equation*}
B^{\prime }:\text{ }x_{0}\text{ }\left[ x_{1}\text{ }x_{2}\right] \,x_{3}%
\text{ }\ldots \text{ }x_{\left\vert X\right\vert -1},
\end{equation*}%
representing that the elements in between $\left[ \text{ \ }\right] $ are
indifferent with $B^{\prime }$.

Let $I$ and $S$ be the sets of students and schools, respectively, where $%
\left\vert I\right\vert \geq 3$. Each student $i\in I$ has a linear order on 
$S\cup \left\{ \emptyset \right\} $ denoted by $P_{i}$ representing the
preference of student $i$, where $sP_{i}s^{\prime }$ (or equivalently $%
\left( s,s^{\prime }\right) \in P_{i}$) means that $i$ prefers $s\in S\cup
\left\{ \emptyset \right\} $ to $s^{\prime }\in S\cup \left\{ \emptyset
\right\} $ and $\emptyset $ represents their best outside option. If $%
sP_{i}\emptyset $, then school $s$ is said to be \textbf{acceptable} for $i$%
. Further, $sR_{i}s^{\prime }$ means $sP_{i}s^{\prime }$ or $s=s^{\prime }$.
Let $\mathbb{P}$ be all possible linear orders on $S\cup \left\{ \emptyset
\right\} $.

Each school $s$ has a capacity constraint represented by $q_{s}\in \mathbb{Z}%
_{++}$ and moreover, $q=\left( q_{s}\right) _{s\in S}.$ Let $\succ _{s}$ be
an asymmetric binary relation on $I$ representing the priority for school $s$%
, where $\left( i,j\right) \in \succ _{s}$ means that $i$ has a higher
priority than $j$ for school $s$. Let $\mathcal{B}$ be the set of all
possible asymmetric binary relations on $I$. Hereafter, we call $\succ
_{s}\in \mathcal{B}$ a \textbf{priority order}.\footnote{%
To put it precisely, $\succ _{s}\in \mathcal{B}$ may be improper to call a
priority \textquotedblleft order\textquotedblright , because it may not be
transitive. However, with few exceptions, we assume that $\succ _{s}$ is
transitive and thus we call it a priority order.} Moreover, let $\mathcal{P}%
, $ $\mathcal{W}$ and $\mathcal{L}$ be the set of all possible partial
orders, weak orders and linear orders on $I$, respectively. Let $\succ
=\left( \succ _{s}\right) _{s\in S}\in \mathcal{B}^{\left\vert S\right\vert
} $ and $\succ ^{\prime }=\left( \succ _{s}^{\prime }\right) _{s\in S}\in 
\mathcal{B}^{\left\vert S\right\vert }$ represent typical profiles of
priority orders on $I$.

For $\succ _{s}\in \mathcal{B},$ let $\succ _{s}^{\ast }$ be a (linear
order) \textbf{extension} of $\succ _{s}$ if $\succ _{s}^{\ast }\in \mathcal{%
L}$ and $\succ _{s}\subseteq \succ _{s}^{\ast }$. This is well-defined
because $\succ _{s}$ is asymmetric. On the existence of extensions of $\succ
_{s}$, the following result is known.

\begin{remark}
There is an extension of $\succ _{s}\in \mathcal{B}$ if and only if $\succ
_{s}$ is acyclic.
\end{remark}

Since any partial order is acyclic, for any $\succ _{s}\in \mathcal{P},$
there is an extension of $\succ _{s}$. Note that there may exist multiple
extensions for $\succ _{s}$. Let $\succ ^{\ast }$ be an \textbf{extension
profile of} $\succ $ if $\succ _{s}^{\ast }$ is an extension of $\succ _{s}$
for all $s\in S$.

We let 
\begin{equation*}
G_{S}=\left( I,S,P,\succ ,q\right)
\end{equation*}%
be a \textbf{school choice problem with a single priority order}. In almost
cases, $\left( I,S,P,q\right) $ is fixed. Thus, the school choice problem is
simply denoted by $G_{S}=\succ $.

A \textbf{matching} $\mu $ is a mapping satisfying $\mu (i)\in S\cup
\{\emptyset \},$ $\mu \left( s\right) \subseteq I$, and $\mu (i)=s$ if and
only if $i\in \mu \left( s\right) $. Note that $\mu (i)=\emptyset $ means
that $i$ is unmatched to any school and $\mu (i)=s\in S$ means that $i$ is
matched to $s$ under a matching $\mu $. A matching $\mu $ is said to be 
\textbf{individually rational} if $\mu \left( i\right) R_{i}\emptyset $ for
all $i\in I$. A matching $\mu $ is said to be \textbf{non-wasteful} if $%
sP_{i}\mu \left( i\right) $ implies $\left\vert \mu \left( s\right)
\right\vert =q_{s}$ for all $i\in I$ and all $s\in S$. A matching $\mu $ 
\textbf{violates} the priority $\succ _{s}$\ of $i\notin \mu \left( s\right) 
$ over $j\in \mu \left( s\right) $ if $sP_{i}\mu \left( i\right) \ $and $%
\left( i,j\right) \in \succ _{s}$. If a matching $\mu $ does not violate the
priority $\succ _{s}$ for any $s\in S$, then it is said to be \textbf{fair }%
for $\succ $. A matching $\mu $ is \textbf{stable} for $\succ $ if it is
individually rational, non-wasteful and fair for $\succ $.

The following result is known (see, e.g., Kitahara and Okumura (2024)).

\begin{remark}
If $\mu $ is stable for $\succ $ and $\succ ^{\prime }$ is an extension
profile of $\succ $, then $\mu $ is also stable for $\succ ^{\prime }$.
\end{remark}

By this result, if there is an extension profile of $\succ $, then a stable
matching for the extension profile is also stable for $\succ $. The
following result is also due to Kitahara and Okumura (2024, Lemma 1).

\begin{remark}
If $\succ _{s}\in \mathcal{B}$ is acyclic for all $s\in S$, then there is a
stable matching for $\succ $.
\end{remark}

By Remark 1, if $\succ _{s}\in \mathcal{B}$ is acyclic, then there is an
extension of $\succ _{s}$ for each $s\in S$. By Remark 2, a stable matching
for the extension profile is also stable for $\succ $. Since Gale and
Shapley (1962) show that there is a stable matching for a profile of linear
priority orders, we have Remark 3.

A matching $\mu $ \textbf{is} \textbf{Pareto dominated for }$I^{\prime }$ 
\textbf{by} $\mu ^{\prime }$ if $\mu ^{\prime }\left( i\right) P_{i}\mu
\left( i\right) $\ for some $i\in I^{\prime }$ and $\mu ^{\prime }\left(
i^{\prime }\right) R_{i^{\prime }}\mu \left( i^{\prime }\right) $ for all $%
i^{\prime }\in I^{\prime }$. Moreover, a matching $\mu $ \textbf{is weakly} 
\textbf{Pareto dominated for }$I^{\prime }$ \textbf{by} $\mu ^{\prime }$ if
it is Pareto dominated for $I^{\prime }$ by $\mu ^{\prime }$ or $\mu \left(
i\right) =\mu ^{\prime }\left( i\right) $ for all $i\in I^{\prime }$. When $%
I^{\prime }=I$ is additionally satisfied, we simply write that $\mu $ 
\textbf{is (weakly) Pareto dominated by} $\mu ^{\prime }$, as defined in
many previous studies. A matching $\mu $ is a \textbf{student optimal stable
matching} (hereafter \textbf{SOSM}) for $\succ $\textbf{\ }if it is stable
for $\succ $ and is not Pareto dominated for all students by any stable
matching for $\succ $.

Second, we consider the case where each school is allowed to have multiple
priority orders on $I$, where each of them is represented by a priority for
the school. Let $\mathcal{\vartriangleright }_{s}\in 2^{\mathcal{B}}$ be the
set of priority orders of school $s$ and $\vartriangleright =\left( \mathcal{%
\vartriangleright }_{s}\right) _{s\in S}$ that is called a \textbf{multiple
priority orders profile}. We let 
\begin{equation*}
G_{M}=\left( I,S,P,\vartriangleright ,q\right)
\end{equation*}%
be a \textbf{school choice problem with multiple priority orders}. Since $%
\left( I,S,P,q\right) $ is fixed, a school choice problem is simply denoted
by $G_{M}=\mathcal{\vartriangleright }$.

We introduce a fairness notion in the case of multiple priority orders. We
consider the situation where $\mu $ violates some priority in $\mathcal{%
\vartriangleright }_{s}$. Then, the violation of the priority $\succ
_{s}\left( \in \mathcal{\vartriangleright }_{s}\right) $ of $i\notin \mu
\left( s\right) $ over $j\in \mu \left( s\right) $ is said to be \textbf{%
justified by} another priority $\succ _{s}^{\prime }\in \mathcal{%
\vartriangleright }_{s}$ if $sP_{i}\mu \left( i\right) ,$ $\left( i,j\right)
\in \succ _{s},$ and $\left( j,i\right) \in \succ _{s}^{\prime }$. In this
case, since a school may have two or more priority orders that are
inconsistent with each other, even if one priority is violated, the
violation can be justified by another priority.

A matching $\mu $ is \textbf{M-fair} for $\vartriangleright $ if $sP_{i}\mu
\left( i\right) $ and $\left( i,j\right) \in \succ _{s}$ for some $\succ
_{s}\in \mathcal{\vartriangleright }_{s}$ imply that there is $\succ
_{s}^{\prime }\in \mathcal{\vartriangleright }_{s}$ such that $\left(
j,i\right) \in \succ _{s}^{\prime }$. That is, if an M-fair matching for $%
\vartriangleright $ violates a priority in $\vartriangleright _{s}$, then it
is justified by another priority in $\vartriangleright _{s}$. On the other
hand, in a matching that is not M-fair for $\vartriangleright $, there is a
school $s$ such that some priority in $\mathcal{\vartriangleright }_{s}$ is
violated and no priority justifies the violation. This implies that in an
M-fair matching, a priority violation is allowed only if it is justified by
some priorities.\footnote{%
We can consider a majority like fairness concept. That is, matching $\mu $
is fair for $\vartriangleright $ if $sP_{i}\mu \left( i\right) $ and $\left(
i,j\right) \in \succ _{s}$ for some $\succ _{s}\in \mathcal{%
\vartriangleright }_{s}$ imply that there are $\left\vert \mathcal{%
\vartriangleright }_{s}\right\vert /2$ or more priority orders $\succ
_{s}^{\prime }\in \mathcal{\vartriangleright }_{s}$ such that $\left(
j,i\right) \in \succ _{s}^{\prime }$. However, in this definition, there may
be no fair matching in the Condorcet cycle case.} Note that if $\mathcal{%
\vartriangleright }_{s}$ is a singleton for all $s\in S$, the definition of
M-fairness is equivalent to that of usual fairness. Moreover, in Appendix
A.1, we introduce another (weaker) fairness notion in the case of multiple
priority orders.

A matching $\mu $ is said to be\textbf{\ M-stable }for $\vartriangleright $
if it is individually rational, non-wasteful and M-fair for $%
\vartriangleright $. A matching is an $I^{\prime }$\textbf{\ optimal M-stable%
} for $\vartriangleright $ if it is M-stable for $\vartriangleright $ and is
not Pareto dominated for $I^{\prime }$ by any M-stable matching for $%
\vartriangleright $. Specifically, when $I^{\prime }=I$, we call $\mu $ a 
\textbf{student-optimal M-stable matching }(hereafter \textbf{SOMSM})\textbf{%
\ }for $\vartriangleright $. By this definition, if there is a matching that
Pareto dominates SOMSM\textbf{\ }for $\vartriangleright $, then there is at
least one priority violation without any justification.

To understand a school choice problem with multiple priority orders, we
introduce an example.

\subsubsection*{Example 1}

First, suppose that there are four students $i_{1},\cdots ,i_{4}$ and one
school $s$. Among them, $i_{1}$ and $i_{2}$ have a sibling who belongs to $s$%
, and $i_{1}$ and $i_{3}$ are living in the school zone of $s$. If $\succ
_{s}$ is a sibling priority order and $\succ _{s}^{\prime }$ is a school
zone priority order, then 
\begin{gather*}
\succ _{s}=\left\{ \left( i_{1},i_{3}\right) ,\left( i_{2},i_{3}\right)
,\left( i_{1},i_{4}\right) ,\left( i_{2},i_{4}\right) \right\} , \\
\succ _{s}^{\prime }=\left\{ \left( i_{1},i_{2}\right) ,\left(
i_{3},i_{2}\right) ,\left( i_{1},i_{4}\right) ,\left( i_{3},i_{4}\right)
\right\} .
\end{gather*}

Suppose $q_{s}=2$ and $sP_{i}\emptyset $ for all $i=i_{1},\cdots ,i_{4}$.
Then, there is no matching that is stable for both $\succ $ and $\succ
^{\prime }$, because $\mu ^{A}$ such that $\mu ^{A}\left( s\right) =\left\{
i_{1},i_{2}\right\} $ violates $\succ _{s}^{\prime }$\ of $i_{3}$ for $s$
over $i_{2}$, and $\mu ^{B}$ such that $\mu ^{B}\left( s\right) =\left\{
i_{1},i_{3}\right\} $ violates $\succ _{s}$\ of $i_{2}$ for $s$ over $i_{3}$%
. However, both $\mu ^{A}$ and $\mu ^{B}$ are M-stable matching\textbf{\ }%
for $\mathcal{\vartriangleright =}\left( \left\{ \succ _{s},\succ
_{s}^{\prime }\right\} \right) $. This is because, since $\left(
i_{2},i_{3}\right) \in \succ _{s}$ and $\left( i_{3},i_{2}\right) \in \succ
_{s}^{\prime }$, the violations mentioned above are allowed.

In a typical school choice system, a linear priority order is established by
determining which priority is more important. For example, in the old system
of Boston, since the sibling priority is considered to be more important
than the school zone priority, the linear order in this example would be%
\begin{equation*}
\succ _{s}^{\ast }:i_{1}\text{ }i_{2}\text{ }i_{3}\text{ }i_{4}\text{.}
\end{equation*}%
See Abdulkadiro\u{g}lu et al. (2006) on the rule. In our approach, we can
attain a matching that weakly Pareto dominates any stable matching for any
profile of linear orders constructed like $\succ _{s}^{\ast }$. To
understand this problem, we extend this example.

We add one school $s^{\prime }$ and, $i_{3}$ and $i_{4}$ have a sibling who
belongs to $s^{\prime }$ and moreover, $i_{2}$ and $i_{4}$ are living in the
school zone of $s^{\prime }$. If $\succ _{s^{\prime }}$ is a sibling
priority order and $\succ _{s^{\prime }}^{\prime }$ is a school zone
priority order, then 
\begin{gather*}
\succ _{s^{\prime }}=\left\{ \left( i_{3},i_{1}\right) ,\left(
i_{4},i_{2}\right) ,\left( i_{3},i_{2}\right) ,\left( i_{4},i_{1}\right)
\right\} , \\
\succ _{s^{\prime }}^{\prime }=\left\{ \left( i_{2},i_{1}\right) ,\left(
i_{4},i_{1}\right) ,\left( i_{2},i_{3}\right) ,\left( i_{4},i_{3}\right)
\right\} .
\end{gather*}%
If the sibling priority is considered to be more important than the school
zone priority (like the case of $\succ _{s}^{\ast }$), then the linear
priority order of $s^{\prime }$ would be 
\begin{equation*}
\succ _{s^{\prime }}^{\ast }:i_{4}\text{ }i_{3}\text{ }i_{2}\text{ }i_{1}%
\text{.}
\end{equation*}

Now, suppose $q_{s}=q_{s^{\prime }}=1$ and 
\begin{equation*}
s^{\prime }P_{1}sP_{1}\emptyset ,\text{ }sP_{2}\emptyset P_{2}s^{\prime },%
\text{ }sP_{3}s^{\prime }P_{3}\emptyset ,\text{ }sP_{4}\emptyset
P_{4}s^{\prime }\text{.}
\end{equation*}%
Then, the unique stable matching for $\succ ^{\ast }=\left( \succ _{s}^{\ast
},\succ _{s^{\prime }}^{\ast }\right) $ is $\mu ^{C}$ such that $\mu
^{C}\left( s\right) =\left\{ i_{1}\right\} $ and $\mu ^{C}\left( s^{\prime
}\right) =\left\{ i_{3}\right\} $. On the other hand, there are two M-stable
matchings\textbf{\ }for $\vartriangleright =\left( \mathcal{%
\vartriangleright }_{s},\mathcal{\vartriangleright }_{s^{\prime }}\right) $ $%
\mu ^{C}$ defined above and $\mu ^{D}$ such that $\mu ^{D}\left( s\right)
=\left\{ i_{3}\right\} $ and $\mu ^{D}\left( s^{\prime }\right) =\left\{
i_{1}\right\} $, where $\mathcal{\vartriangleright }_{s^{\prime }}=\left\{
\succ _{s^{\prime }},\succ _{s^{\prime }}^{\prime }\right\} $. Trivially, $%
\mu ^{D}$ is the unique SOMSM\textbf{\ }for $\vartriangleright $; that is, $%
\mu ^{D}$ Pareto dominates $\mu ^{C}$.\newline

Therefore, there may be an M-stable matching for a multiple priority orders
profile that Pareto dominates any stable matching for a single priority
order profile whose components are linear priority orders that are
established by determining which priority is more important.

We introduce another example.

\subsubsection*{Example 2}

In this example, each school has two priority orders that are determined
based on the scores on two different tests. There are four students $%
i_{1},\cdots ,i_{4}$ and only one school $s$. We assume that all students
want to be assigned to $s$. The scores on the two tests for school $s\in S$
of student $i$ are given by $\sigma _{i}$ and $\sigma _{i}^{\prime }$.
Suppose 
\begin{eqnarray*}
\sigma &=&\left( \sigma _{i_{1}},\cdots ,\sigma _{i_{4}}\right) =\left(
25,40,35,50\right) , \\
\sigma ^{\prime } &=&\left( \sigma _{i_{1}}^{\prime },\cdots ,\sigma
_{i_{4}}^{\prime }\right) =\left( 35,10,45,40\right) .
\end{eqnarray*}%
Then, if $\succ _{s}\in \mathcal{\vartriangleright }_{s}$ and $\succ
_{s}^{\prime }\in \mathcal{\vartriangleright }_{s}$ are respectively
constructed from $\sigma $ and $\sigma ^{\prime }$, then 
\begin{gather*}
\succ _{s}:i_{4}\text{ }i_{2}\text{ }i_{3}\text{ }i_{1}, \\
\succ _{s}^{\prime }:i_{3}\text{ }i_{4}\text{ }i_{1}\text{ }i_{2}.
\end{gather*}

Let $\mu $ be an M-stable matching for $\mathcal{\vartriangleright =}\left(
\left\{ \succ _{s},\succ _{s}^{\prime }\right\} \right) $. If $q_{s}=1$,
then $\mu (s)$ is either $\left\{ i_{4}\right\} $ or $\left\{ i_{3}\right\} $%
, because $\sigma _{i_{3}}<\sigma _{i_{4}}$, $\sigma _{i_{3}}^{\prime
}>\sigma _{i_{4}}^{\prime },$ $\sigma _{i}<\sigma _{i_{4}}$ and $\sigma
_{i}^{\prime }<\sigma _{i_{4}}^{\prime }$ for $i=i_{1}$,$i_{2}$. Similarly,
if $q_{s}=2$, then $\mu (s)$ is either $\left\{ i_{3},i_{4}\right\} $ or $%
\left\{ i_{2},i_{4}\right\} $, and if $q_{s}=3$, then $\mu (s)$ is either $%
\left\{ i_{1},i_{3},i_{4}\right\} $ or $\left\{ i_{2},i_{3},i_{4}\right\} $.

For example, in many markets where schools consider the scores on multiple
tests, the weighted sum of the scores on the tests is used. In this case,
for $\alpha \in \left( 0,1\right) ,$ we let $\succ _{s}^{\alpha }$ be such
that $i\succ _{s}^{\alpha }j$ if and only if 
\begin{equation*}
\alpha \sigma _{i}+\left( 1-\alpha \right) \sigma _{i}^{\prime }>\alpha
\sigma _{j}+\left( 1-\alpha \right) \sigma _{j}^{\prime }.
\end{equation*}%
For example, if $\alpha =0.5$, then $\succ _{s}^{\alpha }:i_{4}$ $i_{3}$ $%
i_{1}$ $i_{2}$. Moreover, in the case where $q_{s}=2$, a stable matching $%
\mu $ for $\succ _{s}^{\alpha }$ must satisfy $\mu (s)=\left\{
i_{3},i_{4}\right\} $, which is also M-stable for $\left\{ \succ _{s},\succ
_{s}^{\prime }\right\} $. This result can be generalized; that is, for any $%
\alpha \in \left( 0,1\right) ,$ a stable matching $\mu $ for $\succ
_{s}^{\alpha }$ must also be M-stable for $\mathcal{\vartriangleright =}%
\left( \left\{ \succ _{s},\succ _{s}^{\prime }\right\} \right) $.

\section{M-stable matchings}

In this section, we characterize the M-stable matchings for $%
\vartriangleright $. Moreover, we discuss the method to derive M-stable
matchings for $\vartriangleright $. To do so, we let $m:2^{\mathcal{B}%
}\rightarrow \mathcal{B}$ be such that $\left( i,j\right) \in m\left( 
\mathcal{\vartriangleright }_{s}\right) $ if and only if $\left( i,j\right)
\in \succ _{s}$ for some $\succ _{s}\in \mathcal{\vartriangleright }_{s}$
and $\left( j,i\right) \notin \succ _{s}^{\prime }$ for all $\succ
_{s}^{\prime }\in \mathcal{\vartriangleright }_{s}$; that is, 
\begin{equation*}
m\left( \mathcal{\vartriangleright }_{s}\right) =\Pi \left(
\bigcup\limits_{\succ _{s}\in \mathcal{\vartriangleright }_{s}}\succ
_{s}\right) .
\end{equation*}%
Then, this is asymmetric; that is, $m\left( \mathcal{\vartriangleright }%
_{s}\right) \in \mathcal{B}$. Moreover, we let 
\begin{equation*}
M\left( \mathcal{\vartriangleright }\right) =\left( m\left( \mathcal{%
\vartriangleright }_{s}\right) \right) _{s\in S}\in \mathcal{B}^{\left\vert
S\right\vert }.
\end{equation*}

To understand this operator, we revisit Example 1. First, 
\begin{eqnarray*}
m\left( \left\{ \succ _{s},\succ _{s}^{\prime }\right\} \right) &=&\left\{
\left( i_{1},i_{2}\right) ,\left( i_{1},i_{3}\right) ,\left(
i_{1},i_{4}\right) ,\left( i_{2},i_{4}\right) ,\left( i_{3},i_{4}\right)
\right\} , \\
m\left( \left\{ \succ _{s^{\prime }},\succ _{s^{\prime }}^{\prime }\right\}
\right) &=&\left\{ \left( i_{4},i_{1}\right) ,\left( i_{4},i_{2}\right)
,\left( i_{4},i_{3}\right) ,\left( i_{2},i_{1}\right) ,\left(
i_{3},i_{1}\right) \right\} .
\end{eqnarray*}%
Neither $\left( i_{3},i_{2}\right) $ nor $\left( i_{2},i_{3}\right) $ is in $%
m\left( \left\{ \succ _{s},\succ _{s}^{\prime }\right\} \right) $ because
both $\left( i_{3},i_{2}\right) \in \succ _{s}$ and $\left(
i_{2},i_{3}\right) \in \succ _{s}^{\prime }$. Moreover, neither $\left(
i_{3},i_{2}\right) $ nor $\left( i_{2},i_{3}\right) $ is in $m\left( \left\{
\succ _{s^{\prime }},\succ _{s^{\prime }}^{\prime }\right\} \right) $. Then,
there are only two stable matchings $\mu ^{C}$ and $\mu ^{D}$ for 
\begin{equation*}
M\left( \mathcal{\vartriangleright }\right) \left( =\left( m\left( \left\{
\succ _{s},\succ _{s}^{\prime }\right\} \right) ,m\left( \left\{ \succ
_{s^{\prime }},\succ _{s^{\prime }}^{\prime }\right\} \right) \right)
\right) .
\end{equation*}%
That is, the set of stable matchings for $M\left( \mathcal{\vartriangleright 
}\right) $ and that of M-stable matchings for $\mathcal{\vartriangleright }$
are equivalent. This fact can be generalized as follows.

\begin{proposition}
A matching $\mu $ is M-stable for $\mathcal{\vartriangleright }$ if and only
if it is stable for $M\left( \mathcal{\vartriangleright }\right) $.
\end{proposition}

\textbf{Proof.} It is sufficient to show that $\mu $ is M-fair for $\mathcal{%
\vartriangleright }$ if and only if it is fair for $M\left( \mathcal{%
\vartriangleright }\right) $.

First, suppose that a matching $\mu $ is M-fair for $\mathcal{%
\vartriangleright }$. We show that $\mu $ is also fair for $M\left( \mathcal{%
\vartriangleright }\right) $. Suppose not; that is, $\mu $ is not fair for $%
M\left( \mathcal{\vartriangleright }\right) $. Then, there is $\left(
i,j,s\right) \in I^{2}\times S$ such that $\mu $ violates $m\left( \mathcal{%
\vartriangleright }_{s}\right) $ of $i$ for $s$ over $j$. Then, $\left(
i,j\right) \in m\left( \mathcal{\vartriangleright }_{s}\right) $ implies $%
\left( i,j\right) \in \succ _{s}$ for some $\succ _{s}\in \mathcal{%
\vartriangleright }_{s}$ and $\left( j,i\right) \notin \succ _{s}^{\prime }$
for all $\succ _{s}^{\prime }\in \mathcal{\vartriangleright }_{s}$,
contradicting the M-fairness of $\mu $ for $\mathcal{\vartriangleright }$.

Second, suppose that a matching $\mu $ is fair for $M\left( \mathcal{%
\vartriangleright }\right) $. We show $\mu $ is M-fair for $\mathcal{%
\vartriangleright }$. Suppose not; that is, $\mu $ is not M-fair for $%
\mathcal{\vartriangleright }$. Then, for $\succ _{s}\in \mathcal{%
\vartriangleright }_{s}$, $\mu $ violates the priority of $i$; that is,
there is $(i,j,\succ _{s})\in I^{2}\times \mathcal{\vartriangleright }_{s}$
such that $sP_{i}\mu \left( i\right) $, $\left( i,j\right) \in \succ _{s}$, $%
j\in \mu \left( s\right) $, without $\succ _{s}^{\prime }\in \mathcal{%
\vartriangleright }_{s}$ such that $\left( j,i\right) \in \succ _{s}^{\prime
}$. Then, we have $\left( i,j\right) \in m\left( \mathcal{\vartriangleright }%
_{s}\right) $. Therefore, for $m\left( \mathcal{\vartriangleright }%
_{s}\right) $, $\mu $ violates the priority of $i\notin \mu \left( s\right) $
for $s$ over $j\in \mu \left( s\right) $, contradicting the stability of $%
\mu $ for $M\left( \mathcal{\vartriangleright }\right) $. \textbf{Q.E.D.}%
\newline

By this equivalence result and Remark 1, we immediately have the following
result.

\begin{corollary}
Let $\succ ^{\ast }$ be an extension profile of $M\left( \mathcal{%
\vartriangleright }\right) $. Then, a stable matching for $\succ ^{\ast }$
is M-stable for $\mathcal{\vartriangleright }$.
\end{corollary}

This result and Remark 2 imply that if each component of $M\left( \mathcal{%
\vartriangleright }\right) $ is acyclic, then we can computationally
efficiently derive an M-stable matching for $\mathcal{\vartriangleright }$
by the following method. First, we construct $M\left( \mathcal{%
\vartriangleright }\right) $. Second, we derive an extension profile of $%
M\left( \mathcal{\vartriangleright }\right) $ denoted by $\succ $.\footnote{%
When $\succ _{s}$ is acyclic, computationally efficient method to derive an
extension of $\succ _{s}$ is known. See, for example, Kitahara and Okumrua
(2023).} Finally, we derive a stable matching for $\succ $ by using an
algorithm such as the DA algorithm (Gale and Shapley (1962)).

To examine an extension profile $\succ ^{\ast }$ of $M\left( \mathcal{%
\vartriangleright }\right) $, we reconsider Example 2. Trivially, for any $%
\alpha \in \left( 0,1\right) ,$ $\succ _{s}^{\alpha }$ is a weak order and
if $\succ _{s}^{\alpha }$ is a linear order, then it is an extension of $%
m\left( \left\{ \succ _{s},\succ _{s}^{\prime }\right\} \right) $. Thus, if
we construct the linear order by using a weighted sum of scores, then it is
an extension of $m\left( \mathcal{\vartriangleright }_{s}\right) $.
Furthermore, in that case, a stable matching for the profile of those
priority orders is M-stable for the profile of sets of priority orders based
on the respective test scores.

However, if\ $M\left( \mathcal{\vartriangleright }\right) $ is not acyclic,
then there may be no stable matching for $M\left( \mathcal{\vartriangleright 
}\right) $ and therefore, there may be no M-stable matching for $\mathcal{%
\vartriangleright }$. Thus, we consider sufficient conditions for the
existence of M-stable matching for $\mathcal{\vartriangleright }$.

\begin{lemma}
If all elements in $\mathcal{\vartriangleright }_{s}$ are weak orders, then $%
m\left( \mathcal{\vartriangleright }_{s}\right) $ is a partial order.
\end{lemma}

\textbf{Proof.} We show that [$\left( i,j\right) \in m\left( \mathcal{%
\vartriangleright }_{s}\right) $ and $\left( j,k\right) \in m\left( \mathcal{%
\vartriangleright }_{s}\right) $] implies $\left( i,k\right) \in m\left( 
\mathcal{\vartriangleright }_{s}\right) $, for all $i,j,k\in I$. Suppose
not; that is, [$\left( i,j\right) \in m\left( \mathcal{\vartriangleright }%
_{s}\right) $ and $\left( j,k\right) \in m\left( \mathcal{\vartriangleright }%
_{s}\right) $] but $\left( i,k\right) \notin m\left( \mathcal{%
\vartriangleright }_{s}\right) $. By $\left( i,k\right) \notin m\left( 
\mathcal{\vartriangleright }_{s}\right) $, there is $\left( k,i\right) \in
\succ _{s}$ for some $\succ _{s}\in \mathcal{\vartriangleright }_{s}$ or $%
\left( i,k\right) \notin \succ _{s}^{\prime }$ for all $\succ _{s}^{\prime
}\in \mathcal{\vartriangleright }_{s}$. Since $\left( i,j\right) \in m\left( 
\mathcal{\vartriangleright }_{s}\right) $ and $\left( j,k\right) \in m\left( 
\mathcal{\vartriangleright }_{s}\right) $, $\left( j,i\right) \notin \succ
_{s}^{\prime }$ and $\left( k,j\right) \notin \succ _{s}^{\prime }$ for all $%
\succ _{s}^{\prime }\in \mathcal{\vartriangleright }_{s}$. Since $\succ
_{s}^{\prime }$ is negatively transitive for all $\succ _{s}^{\prime }\in 
\mathcal{\vartriangleright }_{s}$, $k\not\succ _{s}^{\prime }i$. Therefore, $%
\left( i,k\right) \notin \succ _{s}^{\prime }$ for all $\succ _{s}^{\prime
}\in \mathcal{\vartriangleright }_{s}$. However, $\left( j,i\right) \notin
\succ _{s}^{\prime }$ and $\left( i,k\right) \notin \succ _{s}^{\prime }$
imply $\left( j,k\right) \notin \succ _{s}^{\prime }$ for all $\succ
_{s}^{\prime }\in \mathcal{\vartriangleright }_{s}$. However, this
contradicts $\left( j,k\right) \in m\left( \mathcal{\vartriangleright }%
_{s}\right) $. Therefore, $m\left( \mathcal{\vartriangleright }_{s}\right) $
is transitive. \textbf{Q.E.D.}\linebreak

By Lemma 1 and Remark 3, we have a sufficient condition on $\mathcal{%
\vartriangleright }$ for the existence of M-stable matchings for $\mathcal{%
\vartriangleright }$.

\begin{corollary}
If all elements in $\mathcal{\vartriangleright }_{s}$ are weak orders for
all $s$, then there is a M-stable matching for $\mathcal{\vartriangleright }$%
.
\end{corollary}

\textbf{Proof. }By Remark 3 and Lemma 1, there is a stable matching for $%
M\left( \mathcal{\vartriangleright }\right) $ denoted by $\mu $. By
Proposition 1, $\mu $ is M-stable matching for $\mathcal{\vartriangleright }$%
. \textbf{Q.E.D.}\newline

Note that although acyclicity of $m\left( \mathcal{\vartriangleright }%
_{s}\right) $ for all $s\in S$ is sufficient for the existence of M-stable
matchings for $\mathcal{\vartriangleright }$, the fact that $m\left( 
\mathcal{\vartriangleright }_{s}\right) $ is a partial order (satisfies
transitivity) is important in some results introduced later.

As stated by Erdil and Ergin (2008) and Abdulkadiro\u{g}lu et al. (2009),
the priority orders of schools of several real world markets are weak
orders. By Corollary 2, the existence of M-stable matchings are ensured for
such markets. On the other hand, if $\succ _{s}$ is not a weak order but a
partial order for some $\succ _{s}\in \mathcal{\vartriangleright }_{s}$ and
some $s\in S,$ then $m\left( \mathcal{\vartriangleright }_{s}\right) $ may
not be transitive. Moreover, in that case, $m\left( \mathcal{%
\vartriangleright }_{s}\right) $ may not even be acyclic. We show these
facts.

\subsubsection*{Example 3}

Suppose $I=\left\{ i_{1},i_{2},i_{3}\right\} $ and $S=\left\{ s\right\} $.
First, $\mathcal{\vartriangleright }_{s}=\left\{ \succ _{s}^{1},\succ
_{s}^{2}\right\} $, where 
\begin{equation*}
\succ _{s}^{1}:i_{1},i_{2},i_{3}\text{ and }\succ _{s}^{2}=\left\{ \left(
i_{3},i_{1}\right) \right\} .
\end{equation*}%
Then, $\succ _{s}^{1}$ is a linear order, and $\succ _{s}^{2}$ is a partial
order but not a weak order, because $\left( i_{3},i_{2}\right) \notin \succ
_{s}^{2}$ and $\left( i_{2},i_{1}\right) \notin \succ _{s}^{2},$ but $\left(
i_{3},i_{2}\right) \in \succ _{s}^{2}$. In this case, 
\begin{equation*}
m\left( \mathcal{\vartriangleright }_{s}\right) =\left\{ \left(
i_{1},i_{2}\right) ,\left( i_{2},i_{3}\right) \right\} \text{,}
\end{equation*}%
which is not transitive, because $\left( i_{1},i_{3}\right) \notin m\left( 
\mathcal{\vartriangleright }_{s}\right) $.

In this example, $m\left( \mathcal{\vartriangleright }_{s}\right) $ is
acyclic, because $\left( i_{3},i_{1}\right) \notin m\left( \mathcal{%
\vartriangleright }_{s}\right) $. By Remark 3, there is a stable matching
for $M\left( \mathcal{\vartriangleright }\right) $, which is M-stable for $%
\mathcal{\vartriangleright }$. Thus, this example is one to show that Lemma
1 cannot be generalized. Next, we introduce an example that Corollary 2
cannot be generalized; that is, $m\left( \mathcal{\vartriangleright }%
_{s}\right) $ may not even be acyclic. Suppose $I=\left\{
i_{1},i_{2},i_{3}\right\} $, $S=\left\{ s\right\} $, $sP_{i}\emptyset $ for
all $i\in I$, $q_{s}=1,$ and $\mathcal{\vartriangleright }_{s}^{\prime
}=\left\{ \succ _{s}^{2},\succ _{s}^{3},\succ _{s}^{4}\right\} $, where 
\begin{equation*}
\succ _{s}^{3}=\left\{ \left( i_{1},i_{2}\right) \right\} \text{ and }\succ
_{s}^{4}=\left\{ \left( i_{2},i_{3}\right) \right\} ,
\end{equation*}%
which are also not weak orders but partial orders. Then, 
\begin{equation*}
m\left( \mathcal{\vartriangleright }_{s}\right) =\left\{ \left(
i_{3},i_{1}\right) ,\left( i_{1},i_{2}\right) ,\left( i_{2},i_{3}\right)
\right\}
\end{equation*}%
is not acyclic and trivially there is no M-stable matching for $\mathcal{%
\vartriangleright }$.\newline

Therefore, if some priority orders in $\mathcal{\vartriangleright }_{s}$ are
not weak orders for some $s\in S$, there may be no M-stable matching for $%
\mathcal{\vartriangleright }$. In Section 7, we briefly discuss the case
where priority orders are not weak orders.

\begin{lemma}
Suppose that there is at least one linear order $\succ _{s}^{\prime }\in 
\mathcal{\vartriangleright }_{s}$ for all $s\in S$. Then, $\succ ^{\prime }$
is an extension profile of $M\left( \mathcal{\vartriangleright }\right) $
and thus, $m\left( \mathcal{\vartriangleright }_{s}\right) $ is acyclic for
all $s\in S$. Moreover, a stable matching for $\succ ^{\prime }$ is M-stable
for $\mathcal{\vartriangleright }$.
\end{lemma}

\textbf{Proof.} Let $\succ ^{\prime }$ be such that $\succ _{s}^{\prime }\in 
\mathcal{\vartriangleright }_{s}\cap \mathcal{L}$ for all $s\in S$. We show
that $m\left( \mathcal{\vartriangleright }_{s}\right) \subseteq \succ
_{s}^{\prime }$ for all $s\in S$. Then, $\left( a,b\right) \in \succ
_{s}^{\prime }$ implies $\left( b,a\right) \notin m\left( \mathcal{%
\vartriangleright }_{s}\right) $. Moreover, by the contraposition of this, $%
\left( b,a\right) \in m\left( \mathcal{\vartriangleright }_{s}\right) $
implies $\left( a,b\right) \notin \succ _{s}^{\prime }$. In this case, since 
$\succ _{s}^{\prime }$ is total, $\left( b,a\right) \in \succ _{s}^{\prime }$%
. Hence $m\left( \mathcal{\vartriangleright }_{s}\right) \subseteq \succ
_{s}^{\prime }$ and thus $\succ ^{\prime }$ is an extension profile of $%
M\left( \mathcal{\vartriangleright }\right) $.

Thus, there is an extension of $m\left( \mathcal{\vartriangleright }%
_{s}\right) $.\ By Remark 1, $m\left( \mathcal{\vartriangleright }%
_{s}\right) $ is acyclic for all $s\in S$.

By Remark 2, a stable matching for $\succ ^{\prime }$ is stable for $M\left( 
\mathcal{\vartriangleright }\right) $. By Proposition 1, it is also M-stable
for $\mathcal{\vartriangleright }$. \textbf{Q.E.D.}\newline

By this result, even when some priority orders in $\mathcal{%
\vartriangleright }_{s}$ are not weak orders, if at least one element in $%
\mathcal{\vartriangleright }_{s}$ is a linear order, then a stable matching
for the profile of the linear orders is also M-stable for $\mathcal{%
\vartriangleright }$ and thus the existence of an M-stable matching for $%
\mathcal{\vartriangleright }$ is ensured. Moreover, we have the following
result on SOMSMs for $\mathcal{\vartriangleright }$.

\begin{corollary}
Suppose that there is at least one linear order $\succ _{s}^{\prime }\in 
\mathcal{\vartriangleright }_{s}$ for all $s\in S$. An SOMSM for $\mathcal{%
\vartriangleright }$ is not Pareto dominated by any stable matching for $%
\succ ^{\prime }$.
\end{corollary}

By Lemma 2, a stable matching for $\succ ^{\prime }$ denoted by $\mu $ is
M-stable for $\mathcal{\vartriangleright }$. Thus, by the definition of an
SOMSM, $\mu $ does not Pareto dominate any SOMSMs. On the other hand, if $%
\succ _{s}^{\prime }\in \vartriangleright _{s}$ is not a linear order for
some $s\in S$, then there may exist some stable matching for $\succ ^{\prime
}$ denoted by $\mu ^{\prime }$ that Pareto dominates an SOMSM for $%
\vartriangleright $. However, in that case, $\mu ^{\prime }$ violates a
priority in $\vartriangleright $ and the violation is not justified by any
other priorities in $\vartriangleright $. Therefore, in the remaining
sections, we focus on SOMSMs for $\vartriangleright $.

\section{EADA algorithm}

In this section, we consider a method to derive SOMSMs\textbf{\ }for $%
\vartriangleright $. By Proposition 1, if $\mu $ is an SOSM for $M\left( 
\mathcal{\vartriangleright }\right) $, then it is also an SOMSM\textbf{\ }%
for $\vartriangleright $. Therefore, we introduce a method to derive $\mu $
is an SOSM for a single priority order profile.

We introduce the (simplified) EADA algorithm for $G_{S}=\succ $, which is
introduced by Kesten (2010) and modified by Tang and Yu (2014).\footnote{%
The original EADAM The original EADAM by Kesten (2010), which corresponds to
our EADAM for $\succ _{s}=\emptyset $ for all $s$, aims to achieve an
efficient matching for students. However, here we discuss an efficient
matching under the constraint that certain priorities cannot be violated,
specifically focusing on obtaining the SOSM.
\par
Regarding the modification by Tang and Yu (2014), see footnote 7 of Kitahara
and Okumura (2024).} This is almost the same as that introduced by Tang and
Yu (2014, Subsection 4.2), but they only consider weak priority orders. That
is, we show that the algorithm also attains an SOSM for $\succ $ even when $%
\succ _{s}$ is not weak order but partial order for some $s\in S$.

We roughly introduce the following class of algorithms.

\begin{description}
\item[Round $0$] Let $G_{S}=\succ $ be a school choice problem with a single
priority order profile. Choose $\succ ^{\ast }=\left( \succ _{s}^{\ast
}\right) _{s\in S}\in \mathcal{L}^{\left\vert S\right\vert }$ be an
extension profile of $\succ $.

\item[Round $1$] Run the student-proposing deferred acceptance (hereafter
SPDA) algorithm for $\left( I,S,P,\succ ^{\ast },q\right) $.

\item[Round $k\geq 2$] 

\begin{enumerate}
\item Settle the matching at the underdemanded schools at the resulting
matching of Round $k-1$, and remove these schools and the students either
matched with them or not matched with any schools (matched with the
\textquotedblleft null school\textquotedblright\ in the words of Tang and Yu
(2014)).

\item For each removed student $i$ and each remaining school $s$ that $i$
desires, if there is a remaining student $j$ such that $i\succ _{s}j$, then
remove $s$ from the preference of $j$.

\item Run the SPDA algorithm for the subproblem with only the remaining
schools and students whose preferences may be modified.
\end{enumerate}
\end{description}

The algorithm is stopped when all schools are removed. We formally introduce
the algorithms in Appendix A.2.

The result of the EADA algorithm is dependent on $\succ ^{\ast }$ that is
chosen at Round $0$, because there may exist several extensions of $\succ
_{s}$. Thus, for a problem $G_{S}=\succ $ and an extension profile $\succ
^{\ast }$ of $\succ $, we let $EA\left( \succ ,\succ ^{\ast }\right) 
\mathcal{\ }$be the result of the algorithm introduced above.

Note that Kesten (2010) first considers the EADA algorithm without any
constraints and with an arbitrary linear order profile given. He shows that
the result of the algorithm in this case is efficient (i.e., not Pareto
dominated by any matching). In our terminology, he shows that $EA\left( 
\boldsymbol{\emptyset },\succ ^{\ast }\right) $ is efficient for any linear
order profile $\succ ^{\ast }$, where $\boldsymbol{\emptyset =}\left(
\emptyset ,\cdots ,\emptyset \right) $. He also extends the algorithm to
derive an SOSM when $\succ \in \mathcal{W}^{\left\vert S\right\vert }$.

\begin{theorem}
If $\succ \in \mathcal{P}^{\left\vert S\right\vert }$ and $\succ ^{\ast }$
is an extension profile of $\succ $, then $EA\left( \succ ,\succ ^{\ast
}\right) $ is an SOSM for $\succ $.
\end{theorem}

The proof is provided in Appendix A.3.

This result is a generalization of Proposition 5 of Kesten (2010) that is
modified in 4.2 of Tang and Yu (2014); that is, the EADA algorithm attains
an SOSM for $\succ $ even if $\succ _{s}$ is not weak order (but a partial
order) for some $s\in S$. Moreover, by using Theorem 1, we can confirm that
the aforementioned result in Kesten (2010) holds; that is, $EA\left( 
\boldsymbol{\emptyset },\succ ^{\ast }\right) $ is efficient for any linear
order profile $\succ ^{\ast }$, because $\succ ^{\ast }$ must be an
extension profile of $\succ =\boldsymbol{\emptyset }$.

The transitivity of $\succ _{s}$ for all $s\in S$ is important for Theorem
1; that is, if $\succ _{s}\in \mathcal{B\setminus P}$ for some $s\in S$,
then $EA\left( \succ ,\succ ^{\ast }\right) $ may not be stable for $\succ $
even when $\succ _{s}$ is acyclic and $\succ ^{\ast }$ is its extension
profile. We show this fact by providing an example in Appendix A.4.

Although Theorem 1 is a result of a school choice problem with a single
priority order, we utilize this for the analysis of that with multiple
priority orders. With multiple priority orders $\vartriangleright $, we
immediately have the following result.

\begin{corollary}
If $\succ _{s}\in \mathcal{W}$ for all $\succ _{s}\in \vartriangleright _{s}$
and $\succ ^{\ast }$ is an extension profile of $M\left( \mathcal{%
\vartriangleright }\right) $, then $EA\left( M\left( \mathcal{%
\vartriangleright }\right) ,\succ ^{\ast }\right) $ is an SOMSM for $%
\vartriangleright $ and weakly Pareto dominates any stable matching for $%
\succ ^{\ast }$.
\end{corollary}

By Lemma 1, in this case, $m\left( \mathcal{\vartriangleright }_{s}\right) $
is a partial order for all $s\in S$. By Theorem 1, $EA\left( M\left( 
\mathcal{\vartriangleright }\right) ,\succ ^{\ast }\right) $ is an SOSM for $%
M\left( \mathcal{\vartriangleright }\right) $. Therefore, by Proposition 1,
it is an SOMSM for $\vartriangleright $. Note that if $M\left( \mathcal{%
\vartriangleright }\right) $ is not partial order, then it may not be an
SOMSM for $\vartriangleright $. On the other hand, if all elements in $%
\mathcal{\vartriangleright }_{s}$ are weak orders for all $s\in S$, then $%
EA\left( M\left( \mathcal{\vartriangleright }\right) ,\succ ^{\ast }\right) $
must be an SOMSM for $\vartriangleright $.

Moreover, by the construction of the EADA algorithms, the SPDA result of
Round 1 is an SOSM for $\succ ^{\ast }$ that Pareto dominates each of the
other stable matchings for $\succ ^{\ast }$. Since $EA\left( M\left( 
\mathcal{\vartriangleright }\right) ,\succ ^{\ast }\right) $ must weakly
Pareto dominate the SPDA result of Round 1, $EA\left( M\left( \mathcal{%
\vartriangleright }\right) ,\succ ^{\ast }\right) $ weakly Pareto dominates
any stable matching for $\succ ^{\ast }$.

We note the following two facts on an extension profile of $M\left( \mathcal{%
\vartriangleright }\right) $. First, as also considered in Example 2, we
consider the situation where each element of $\mathcal{\vartriangleright }%
_{s}$ is derived from the scores of all students on a different test. Let $%
\succ _{s}^{\ast }$ ($\succ _{s}^{\alpha }$ when there are only two priority
orders) be a linear order derived from the weighted sum of the scores of
each student on the tests, where the weight of each test is positive. Then, $%
\succ _{s}^{\ast }$ must be a linear order extension of $m\left( \mathcal{%
\vartriangleright }_{s}\right) $. Corollary 4 implies that any stable
matching for $\succ ^{\ast }$, which is derived using any weights, is weakly
Pareto dominated by $EA\left( M\left( \mathcal{\vartriangleright }\right)
,\succ ^{\ast }\right) $.\footnote{%
However, unlike the unconstrained EADAM $EA\left( \emptyset ,\succ ^{\ast
}\right) $, $EA\left( M\left( \mathcal{\vartriangleright }\right) ,\succ
^{\ast }\right) $ is an efficient matching under the constraint that school $%
s$ prioritizes $i$ over $j$ when the scores of $s$ on each test are not
lower than those of $j$, and the score of $i$ on at least one test is higher
than that of $j$.}

Second, we consider the situation where at least one linear order $\succ
_{s}^{\prime }$\ exists in $\mathcal{\vartriangleright }_{s}$. Then, by
Lemma 2, $\succ _{s}^{\prime }$ is used as an extension of $m\left( \mathcal{%
\vartriangleright }_{s}\right) $. Therefore, if all elements in $\mathcal{%
\vartriangleright }_{s}$ are weak orders and there is a linear order $\succ
_{s}^{\prime }\in \mathcal{\vartriangleright }_{s}$ for all $s\in S$, then $%
EA\left( M\left( \mathcal{\vartriangleright }\right) ,\succ ^{\prime
}\right) $ must weakly Pareto dominate any stable matching for that linear
order priority profile.\footnote{%
Note that an existence of a linear order in $\mathcal{\vartriangleright }%
_{s} $ for all $s\in S$ is not sufficient for Corollary 3, because in that
case, $m\left( \mathcal{\vartriangleright }_{s}\right) $ is acyclic but may
not be a partial order. See the former example provided in Example 1. Thus,
the condition that all elements in $\mathcal{\vartriangleright }_{s}$ are
weak orders is also important.}

In many school choice systems, a single priority order profile, whose
components are linear orders derived in the manner described above, is
commonly used. By Corollary 4, we can (weakly) improve the result by using
our mechanism. On the other hand, suppose that there is a weak order in $%
\mathcal{\vartriangleright }_{s}$ for some $s$ and we focus on $\succ
^{\prime \prime }$ such that $\succ _{s}^{\prime \prime }\in \mathcal{%
\vartriangleright }_{s}\cap \mathcal{L}$ for some $s$. In this case, a
result of this mechanism may be Pareto dominated by a stable matching for $%
\succ ^{\prime \prime }$ denoted by $\mu ^{\prime }$. However, as shown in
Proposition 1, in $\mu ^{\prime }$, some priority is violated without any
justification of other priorities.

We can derive an SOMSM for $\vartriangleright $ by using an algorithm other
than the EADA algorithm. Kitahara and Okumura (2021) show that if $\succ
_{s} $ is a partial order for all $s\in S$, the stable improvement cycle
algorithms \`{a} la Erdil and Ergin (2008) with a modification also attain
an SOSM for $\succ $ that weakly Pareto dominates any stable matchings for $%
\succ ^{\prime }$, which is an arbitrary profile of linear orders in $%
\vartriangleright _{s}$ for all $s\in S$. By Lemma 1, an SOMSM for $%
\vartriangleright $ that weakly Pareto dominates any stable matching for $%
\succ ^{\prime }$ can also be obtained by utilizing the algorithm instead of
the EADA algorithm. However, in Section 6, we show that the mechanism with
the EADA algorithm has other remarkable properties.

\section{Improvements}

In this section, we specifically consider the case where each school has
only two linear priority orders, which are given by $\succ =\left( \succ
_{s}\right) _{s\in S}\in \mathcal{L}^{\left\vert S\right\vert }$ and $\succ
^{\prime }=\left( \succ _{s}^{\prime }\right) _{s\in S}\in \mathcal{L}%
^{\left\vert S\right\vert }$. We say that $\succ ^{\prime }$ is an \textbf{%
improvement} of $\succ $ for $I^{\prime }\subseteq I$ if for all $i\in I$,
if there is some $j\in I$ and some $s\in S$ such that $\left( j,i\right) \in
\succ _{s}$ and $\left( i,j\right) \in \succ _{s}^{\prime }$, then $i\in
I^{\prime }$.\footnote{%
As shown in Appendix A.5, this definition of improvement is weaker than that
of Jiao and Tian (2019), Jiao and Shen (2021) and Jiao et al. (2022).} Note
that in this definition, if $\succ ^{\prime }$ is an improvement of $\succ $
for $I^{\prime }$ and $I^{\prime }\subseteq I^{\prime \prime }\subseteq I$,
then $\succ ^{\prime }$ is also an improvement of $\succ $ for $I^{\prime
\prime }$. Moreover, for any $\succ ,$ $\succ ^{\prime }=\left( \succ
_{s}^{\prime }\right) _{s\in S}\in \mathcal{L}^{\left\vert S\right\vert }$, $%
\succ ^{\prime }$ is an improvement of $\succ $ (also $\succ $ is an
improvement of $\succ ^{\prime }$)$\ $for the set of all students $I$.
Therefore, if we consider a sufficiently large set $I^{\prime }$, any
priority profile $\succ $ is an improvement of any priority profile $\succ
^{\prime }$. However, if the set is small, the priority orders with an
improvement relationship are limited.

As an example, we consider the priority-based affirmative action policy.
Suppose that the students are divided into the majority students and the
minority students. Let $\succ $ be the profile of original priority orders
that determined by the test scores of students and $\succ ^{\prime }$ be the
profile of priority orders that are adjusted by a priority-based affirmative
action policy. Since the policy is aim to create more opportunities for
minority students, $\succ ^{\prime }$ should be an improvement of $\succ $
for the minority students.

We provide an example to understand the definitions.

\subsubsection*{Example 4}

Let $I=\left\{ i_{1},\cdots ,i_{6}\right\} $ and $S=\left\{ s\right\} $.
Here, we assume that $i_{1},\cdots ,i_{4}$ are minority students and $%
i_{5},i_{6}$ are majority students. The score on the entrance exam for
school $s\in S$ of student $i$ is given by $\sigma _{i}$. Suppose 
\begin{equation*}
\left( \sigma _{i_{1}},\cdots ,\sigma _{i_{6}}\right) =\left(
611,656,632,660,643,705\right) .
\end{equation*}%
First, we let $\succ _{s}^{1}$ be the priority order that simply determined
by the scores; that is, 
\begin{equation*}
\succ _{s}^{1}:\text{ }i_{6}\text{ }i_{4}\text{ }i_{2}\text{ }i_{5}\text{ }%
i_{3}\text{ }i_{1}.
\end{equation*}%
On the other hand, we consider some bonus points for the minority students.
Let $\alpha _{i}$ (also $\alpha _{i}^{\prime }$ and $\alpha _{i}^{\prime
\prime }$) be the bonus point for student $i$ and 
\begin{eqnarray*}
\alpha &=&\left( \alpha _{i_{1}},\cdots ,\alpha _{i_{6}}\right) =\left(
50,50,10,10,0,0\right) , \\
\alpha ^{\prime } &=&\left( \alpha _{i_{1}}^{\prime },\cdots ,\alpha
_{i_{6}}^{\prime }\right) =\left( 50,50,30,30,0,0\right) , \\
\alpha ^{\prime \prime } &=&\left( \alpha _{i_{1}}^{\prime \prime },\cdots
,\alpha _{i_{6}}^{\prime \prime }\right) =\left( 60,60,10,10,0,0\right) .
\end{eqnarray*}%
With the bonus points profile $\alpha ,\alpha ^{\prime }$ and $\alpha
^{\prime \prime }$, the priority order $\succ _{s}^{2},\succ _{s}^{3}$ and $%
\succ _{s}^{4}$ are respectively such that%
\begin{gather*}
\succ _{s}^{2}:\text{ }i_{2}\text{ }i_{6}\text{ }i_{4}\text{ }i_{1}\text{ }%
i_{5}\text{ }i_{3}, \\
\succ _{s}^{3}:\text{ }i_{2}\text{ }i_{6}\text{ }i_{4}\text{ }i_{3}\text{ }%
i_{1}\text{ }i_{5}, \\
\succ _{s}^{4}:\text{ }i_{2}\text{ }i_{6}\text{ }i_{1}\text{ }i_{4}\text{ }%
i_{5}\text{ }i_{3}.
\end{gather*}%
That is, for example, $i\succ _{s}^{2}j$ if and only if 
\begin{equation*}
\alpha _{i}+\sigma _{i}>\alpha _{j}+\sigma _{j}\text{.}
\end{equation*}%
Then, $\succ ^{2}$, $\succ ^{3}$ and $\succ ^{4}$ are improvements of $\succ
^{1}$ for $\left\{ i_{1},\cdots ,i_{4}\right\} $.\footnote{%
By our definition, $\succ ^{2}$ and $\succ ^{4}$ are improvements of $\succ
^{1}$ for any superset of $\{i_{1},i_{2}\}$, and $\succ ^{3}$ is improvement
of $\succ ^{1}$ for any superset of $\{i_{1},i_{2},i_{3}\}$.}

As explained above, if a sufficiently large subset of students is regarded
as the improved group, then any priority order can be seen as an improvement
of any other priority order. Thus, the notable point here is that in these
cases, it is not necessary to include the majority of students in the
improved group, because the bonus points for each majority student are zero. 
\newline

Hereafter, in this section, we fix the profile of original priority orders $%
\succ $ and the improved students $I^{\prime }\subseteq I,$ and let 
\begin{eqnarray*}
\vartriangleright _{s}^{\prime } &=&\left\{ \succ _{s},\succ _{s}^{\prime
}\right\} ,\text{ }\vartriangleright _{s}^{\prime \prime }=\left\{ \succ
_{s},\succ _{s}^{\prime \prime }\right\} , \\
\vartriangleright ^{\prime } &=&\left( \vartriangleright _{s}^{\prime
}\right) _{s\in S},\text{ }\vartriangleright ^{\prime \prime }=\left(
\vartriangleright _{s}^{\prime \prime }\right) _{s\in S},
\end{eqnarray*}%
where $\succ ^{\prime }$ and $\succ ^{\prime \prime }$ are improvements of $%
\succ $ for $I^{\prime }\subseteq I$.

Then, we have the following result.

\begin{theorem}
Suppose that $\succ ^{\prime }\in \mathcal{L}$ is an improvement of $\succ
\in \mathcal{L}$ for $I^{\prime }$. Then, $EA\left( M\left( \mathcal{%
\vartriangleright }^{\prime }\right) ,\succ ^{\prime }\right) $ is an $%
I^{\prime }$\textbf{\ }optimal M-stable matching for $\vartriangleright
^{\prime }$.
\end{theorem}

The proof is provided in Appendix A.6.

Theorem 2 implies that if each school has two priority orders, and one is an
improvement for a group of students over the other, then the EADA algorithm
results in an improved-group optimal M-stable matching for the profile of
the two priority orders of all schools. In the case of priority-based
affirmative action, this mechanism achieves minority optimality. That is,
even if the welfare of the majority students can be reduced by any amount as
long as the matching is M-stable, the result of the EADA algorithm is one
where no minority student can be made better off without making some other
minority student worse off.

Of course, this does not mean that all members of an improved group will
never be worse off, because there may be competition among the members.
However, if some members of the group are worse off, then there must exist
other members who are better off.

Next, we consider two improvements of $\succ $ denoted by $\succ ^{\prime }$
and $\succ ^{\prime \prime }$ for a group of students $I^{\prime }$. We say
that $\succ ^{\prime }$ \textbf{more improves} $\succ $ \textbf{than} $\succ
^{\prime \prime }$ for $I^{\prime }\subseteq I$ if

\begin{description}
\item[(i)] $\succ ^{\prime }$ and $\succ ^{\prime \prime }$ are improvements
of $\succ $ for $I^{\prime },$

\item[(ii)] $\succ ^{\prime }$ is an improvement of $\succ ^{\prime \prime }$
for $I^{\prime }$, and

\item[(iii)] for all $s\in S$ and all $i,i^{\prime }\in I$, $\left(
i,i^{\prime }\right) \in \succ _{s}$ and $\left( i^{\prime },i\right) \in
\succ _{s}^{\prime \prime }$ imply $\left( i^{\prime },i\right) \in \succ
_{s}^{\prime }$.\footnote{%
If (i) and (ii) are satisfied, then $i^{\prime }\in I^{\prime }$.}
\end{description}

We consider Condition (iii). Suppose $\left( i,i^{\prime }\right) \in \succ
_{s}$, $\left( i^{\prime },i\right) \in \succ _{s}^{\prime \prime },$ and
Conditions (i) and (ii) are satisfied. Since $\succ ^{\prime \prime }$ is an
improvement of $\succ $ for $I^{\prime }$, $i^{\prime }\in I^{\prime }$.
Then, $\left( i,i^{\prime }\right) \in \succ _{s}^{\prime }$ is allowed to
be satisfied, when $i\in I^{\prime }$ is also satisfied. However, Condition
(iii) does not allow it; that is, if $i^{\prime }$ has a priority over $i$
under $\succ _{s}^{\prime \prime }$, then $i^{\prime }$ also has the
priority under the more improved priority order $\succ _{s}^{\prime }$; that
is, the improvement is monotone to the improved group. Condition (iii) is
important for the results introduced later in this section. We show the
importance of the condition on them in Appendix A.5.

To understand this definition, we revisit Example 3. Then, $\succ ^{4}$ more
improves $\succ ^{1}$ than $\succ ^{2}$ for $\left\{ i_{1},\cdots
,i_{4}\right\} $. However, $\succ ^{3}$ does not more improves $\succ ^{1}$
than $\succ ^{2}$ for $\left\{ i_{1},\cdots ,i_{4}\right\} $, because $%
\left( i_{3},i_{1}\right) \in \succ _{s}^{1}$, $\left( i_{1},i_{3}\right)
\in \succ _{s}^{2}$ but $\left( i_{3},i_{1}\right) \in \succ _{s}^{3}$; that
is, Condition (iii) is not satisfied. In this example, the score of $i_{3}$
is 21 points higher than that of $i_{1}$. With the bonus points profile $%
\alpha $, this is reversed, because the bonus point of $i_{1}$ is 40 points
higher than that of $i_{3}$. On the other hand, with the bonus points
profile $\alpha ^{\prime }$, this reversal does not occur, because the bonus
point of $i_{1}$ is only 20 points higher than that of $i_{3}$. Therefore,
in cases where there is a change in the priority among minorities, the
statement with "more improves" is not satisfied.

Now, we explicitly define mechanisms. A mechanism $\phi :\mathcal{L}%
^{\left\vert S\right\vert }\times \mathcal{L}^{\left\vert S\right\vert
}\rightarrow \mathcal{M}$ is a mapping where $\mathcal{M}$ is the set of all
matchings. A mechanism $\phi $ is said to be \textbf{stable }if for any $%
\succ ^{\prime }\in \mathcal{L}^{\left\vert S\right\vert },$ $\phi \left(
\succ ,\succ ^{\prime }\right) $ is stable for $\succ ^{\prime }$. Note that
this stability notion only cares one of the two priority order profiles.

A mechanism $\phi $ is said to be \textbf{responsive to improvements} if the
fact that $\succ ^{\prime }\in \mathcal{L}^{\left\vert S\right\vert }$ more
improves $\succ \in \mathcal{L}^{\left\vert S\right\vert }$ than $\succ
^{\prime \prime }\in \mathcal{L}^{\left\vert S\right\vert }$ for $I^{\prime
}\subseteq I$ implies that $\phi \left( \succ ,\succ ^{\prime }\right) $ is
not Pareto dominated for $I^{\prime }$ by $\phi \left( \succ ,\succ ^{\prime
\prime }\right) $. As mentioned above, the definition of improvement in this
study is weaker than that in several previous studies. Thus, the
responsiveness property is stronger than that considered in previous
studies, such as Jiao and Shen (2021) and Jiao et al. (2022). We show this
fact in Appendix A.5.

The following impossibility result is straightforward from a result of
Kojima (2012, Theorem 2).

\begin{remark}
There is no mechanism that is stable and responsive to improvements.
\end{remark}

By this impossibility result, our attention shifts to an M-stable mechanism.
Thus, we introduce two properties of the mechanisms.

A mechanism $\phi $ is said to be a \textbf{student} \textbf{optimally} 
\textbf{M-stable} if for any $\succ ,\succ ^{\prime }\in \mathcal{L}%
^{\left\vert S\right\vert },$ $\phi \left( \succ ,\succ ^{\prime }\right) $
is an SOMSM for $\vartriangleright ^{\prime }\left( =\left( \left\{ \succ
_{s},\succ _{s}^{\prime }\right\} _{s\in S}\right) \right) $. Moreover, a
mechanism $\phi $ is said to be an \textbf{improved-group} \textbf{optimally}
\textbf{M-stable} if for any $\succ ,\succ ^{\prime }\in \mathcal{L}%
^{\left\vert S\right\vert }$ such that $\succ ^{\prime }$ is an improvement
for $I^{\prime }\subseteq I$, $\phi \left( \succ ,\succ ^{\prime }\right) $
is an $I^{\prime }$ optimal M-stable matching for $\vartriangleright
^{\prime }$. Note that since $\succ ^{\prime }$ is an improvement of $\succ
\ $for the set of all students $I$, an improved-group optimally M-stable
mechanism is also student optimally M-stable.

Now, we introduce a mechanism%
\begin{equation*}
\phi ^{\ast }\left( \succ ,\succ ^{\prime }\right) =EA\left( M\left( 
\mathcal{\vartriangleright }^{\prime }\right) ,\succ ^{\prime }\right) \text{%
.}
\end{equation*}%
We show that $\phi ^{\ast }$ is responsive to improvements.

\begin{theorem}
The mechanism $\phi ^{\ast }$ is improved-group optimally M-stable and
responsive to improvements.
\end{theorem}

We introduce the proof of Theorem 3 in Appendix A.7.

By Remark 4, there is no stable mechanism that is responsive to
improvements. On the other hand, if we consider M-stability instead of usual
stability, then an M-stable mechanism $\phi ^{\ast }$ is responsive to
improvements. Moreover, it always attains an improved-group (as well as
whole students) optimally M-stable matching. Thus, using this mechanism is
consistent with the policy aim of providing better educational opportunities
for minority students.

We can consider some other definitions of improvements and the
responsiveness property. In Appendix A.5, we introduce a stronger property,
but the responsiveness in that definition is shown to be incompatible with
M-stablility.

Furthermore, we show that $\phi ^{\ast }$ is responsive to improvements
within the domain $\mathcal{L}^{\left\vert S\right\vert }\times \mathcal{L}%
^{\left\vert S\right\vert }$. If we consider a wider domain $\mathcal{W}%
^{\left\vert S\right\vert }\times \mathcal{W}^{\left\vert S\right\vert }$,
then $\phi ^{\ast }$ is not responsive to improvements. In Appendix A.8, we
provide an example to show this fact.

\section{Other applications}

The following are possible applications on our results.

\begin{description}
\item[Multiple affirmative action policies] According to S\"{o}nmez and
Yenmez (2022), in school choice matching markets in several countries such
as Chile, India, and Brazil, there are multiple affirmative action policies,
which implies that there are multiple target beneficiary groups. In these
markets, there are students who belong to two or more of these groups,
resulting in overlapping groups. When priority-based affirmative action
policies are established for $n$ target beneficiary groups, we have $n$
priority orders. By using the EADA algorithm and these priority orders, we
have an SOMSM for the profile of the sets of priority orders.

\item[Chinese college admission] According to Fang and Yasuda (2021), in the
Chinese college admission market, test scores and colleges' preferences
coexist as two possible priority orders for one school. Let $%
\vartriangleright _{s}$ be the set of the priority orders based on the score
and the preference of $s$. In this case, an M-stable matching for $%
\vartriangleright $ is immune to priority violations with respect to both
priority orders. For example, under an M-stable matching for $%
\vartriangleright $, if the test score of student $i$ who wishes to be
matched to $s$ is higher than student $j$ who is matched to $s$, then school
(college) $s$ prefers $j$ to $i$. In the Appendix A.9., we explain the
difference of our analysis and that of Fang and Yasuda (2021). We achieve
the Pareto improvement the matching by using the EADA algorithm and the two
priority orders.

\item[Merger of several districts] Suppose that there are several districts,
each with a school choice program that may differ from the others. As
discussed by Ortega (2018), when several districts merge into one, it is
inevitable that some agents will become worse off if the matching mechanism
used is stable. In the premerger markets, each district may have a law
and/or government policy to determine the priority orders of schools.
Therefore, a school may have multiple priority orders based on the criterion
of the districts. By using our mechanism and these priority orders of all
schools, we achieve a matching that Pareto dominates that obtained in the
premerger markets.

\item[Concentual] Suppose that there are several referees in each school.
Here, we assume that blocking by a school-student pair is established if all
referees of the school does not disagree and at least one of them agrees the
blocking. Let $\vartriangleright _{s}$ be the set of preference relations of
the referees in $s$. An M-stable (M\textbf{-}fair) matching for $%
\vartriangleright $ is immune from any of such blockings.
\end{description}

Kitahara and Okumura (2021) mention that for some markets, the priorities
for schools should be represented by partial orders. For example, consider
the priority order is determined by the distance or the score. We would like
to ignore the negligible difference (say $100$ meters or $1$ point) to
improve the welfare of students. However, if the priority order is required
to be weak, then letting each pair of students with a negligible score
difference being indifferent implies all students being indifferent, because
of the negative transitivity. Therefore, ignoring a small difference and
respecting a large difference are not compatible under weak ordering, but
they are compatible under partial ordering.

Unfortunately, if we use such a priority order as an element of $%
\vartriangleright _{s}$, then $EA\left( M\left( \vartriangleright \right)
,\succ ^{\prime }\right) $ may not be M-stable for $\vartriangleright $.
First, as shown in Example 2, if an element of $\vartriangleright _{s}$ is a
partial order, then $m\left( \vartriangleright _{s}\right) $ may not be
partial order. Second, as shown in Appendix A.4, in the case where $\succ
_{s}$ is not partial order for some $s\in S$, $EA\left( \succ ,\succ ^{\ast
}\right) $ may not be stable for $\succ $, even if $\succ _{s}^{\ast }$ is
an extension of $\succ _{s}$ for all $s\in S$. Therefore, in that case, $%
EA\left( M\left( \vartriangleright \right) ,\succ ^{\prime }\right) $ is not
necessarily M-stable for $\vartriangleright $. In Appendix A.1, we try to
solve this problem by further weakening stability notion.

Finally, in this study, we consider a weaker requirement of stability than
usual. One reason for this is that there may be no non-wasteful matching
that respects all priority orders. However, if non-wastefulness is not
required, a matching that respects all priority orders can exist. We plan to
conduct future research on matchings that respect all priority orders and
are not Pareto dominated by any other matchings that also respect all
priority orders. Kojima and Kamada (2024) also focus on such matchings,
called student optimal fair matchings.

\section*{References}

\begin{description}
\item Abdulkadiro\u{g}lu, A., Che, Y.-K. Pathak, P.A., Roth, A.E., Tercieux,
O. 2020. Efficiency, justified envy, and incentives in priority-based
matching. American Economic Review Insights 2, 425--442.

\item Abdulkadiro\u{g}lu, A., Grigoryan, A., 2024. Diversity Balance in
Centralized Public School Admissions, AEA Papers and Proceedings 114,
497-501.

\item Abdulkadiro\u{g}lu, A., Pathak, P.A., Roth, A.E., 2009.
Strategy-proofness versus efficiency in matching with indifferences:
redesigning the NYC high school match.\ American Economic Review 99,
1954--1978.

\item Abdulkadiro\u{g}lu, A., Pathak, P.A., Roth, A.E., S\"{o}nmez, T. 2005.
The Boston Public School Match.\ American Economic Review 95(2), 368-371.

\item Abdulkadiro\u{g}lu, A., Pathak, P.A., Roth, A.E., S\"{o}nmez, T. 2006.
Changing the Boston Public School Mechanism: Strategy-proofness as equal
access,\ NBER Working paper, 11965.

\item Abdulkadiro\u{g}lu A., S\"{o}nmez, T. 2003. School choice: A mechanism
design approach. American Economic Review 93(3), 729--747.

\item Afacan, M.O., Salman, U. 2016. Affirmative actions: The Boston
mechanism case. Economics Letters 141, 95--97.

\item Alcalde, J., Romero-Medina, A. 2017. Fair student placement. Theory
and Decision 83, 293--307.

\item Alva, S., Manjunath, V. 2019. \textquotedblleft Strategy-proof
Pareto-improvement.\textquotedblright\ Journal of Economic Theory 181,
121--142.

\item Aziz, H., Bir\'{o}, P., Gaspers, S. de Haan, R., Mattei, N.,
Rastegari, B. 2020. Stable matching with uncertain linear preferences.
Algorithmica 82(5), 1410--1433.

\item Balinski, M., S\"{o}nmez, T. 1999. A tale of two mechanisms: student
placement,\ Journal of Economic Theory 84(1), 73--94.

\item Cerrone, C., Hermstr\"{u}wer, Y., Kesten, O. 2024. School Choice with
Consent: An Experiment, Economic Journal 134(661), 1760--1805.

\item Che, Y-K., Kim, J., Kojima, F. 2019a. Stable matching in large
economies. Econometrica, 65-110.

\item Che, Y-K., Kim, J., Kojima, F. 2019b. Weak monotone comparative
statics, Mimeo Available at arXiv:1911.06442

\item Chen, J., Niedermeier, R., Skowron, P. 2018. Stable marriage with
multi-modal preferences. In Proceedings of the 19th ACM Conference on
Economics and Computation. 269--286.

\item Do\u{g}an B. 2016. Responsive affirmative action in school choice.
Journal of Economic Theory 165, 69--105

\item Do\u{g}an B., Ehlers, L. 2021. Minimally unstable Pareto improvements
over deferred acceptance. Theoretical Economics 16, 1249--1279

\item Duggan, J. 1999. A General Extension Theorem for Binary Relations,
Journal of Economic Theory 86, 1-16.

\item Dur, U., Gitmez, A., Y\i lmaz, \"{O}. 2019. School choice under
partial fairness,\ Theoretical Economics 14(4), 1309-1346.

\item Dur, U., Kominers, S.D., Pathak, P.A., S\"{o}nmez, T. 2018. Reserve
Design: Unintended Consequences and the Demise of Boston's Walk Zones.
Journal of Political Economy 126(6) 2457--2479

\item Dur, U., Xie, Y. 2023. Responsiveness to priority-based affirmative
action policy in school choice. Journal of Public Economic Theory 25(2),
229-244

\item Ehlers, L., Morrill, T. 2020. (Il)legal assignments in school choice.
Review of Economic Studies 87, 1837--1875.

\item Erdil, A., Ergin, H. 2008. What's the matter with tie-breaking?
Improving efficiency in school choice,\ American Economic Review 98(3),
669--689.

\item Fang, Y., Yasuda, Y. 2021. Misalignment between Test Scores and
Colleges' Preferences: Chinese College Admission Reconsidered. Available at
SSRN: https://ssrn.com/abstract=3914742

\item Gale D., Shapley L.S. 1962. College admissions and the stability of
marriage. American Mathematical Monthly 69(1):9--15

\item Hafalir I.E., Yenmez M.B., Yildirim M.A. 2013. Effective affirmative
action in school choice. Theoretical Economics 8(2):325--363

\item Hirata, D., Kasuya, Y., Okumura, Y. 2022. Stability,
Strategy-Proofness, and Respect for Improvements, Mimeo Available at SSRN:
https://ssrn.com/abstract=3876865

\item Jiao, Z., Shen, Z. 2021. School choice with priority-based affirmative
action: A responsive solution, Journal of Mathematical Economics 92, 1--9.

\item Jiao, Z., Shen, Z., Tian, G. 2022. When is the deferred acceptance
mechanism responsive to priority-based affirmative action? Social Choice and
Welfare 58, 257--282.

\item Jiao, Z., Tian, G., 2018. Two further impossibility results on
responsive affirmative action in school choice. Economics Letters 166,
60--62.

\item Kamada, Y., Kojima, F. 2024. Fair Matching under Constraints: Theory
and Applications, Review of Economic Studies 91(2), 1162--1199.

\item Kesten, O. 2010. School choice with consent. Quarterly Journal of
Economics 125(3), 1297--1348.

\item Kitahara, M., Okumura, Y. 2020. Stable Improvement Cycles in a
Controlled School Choice. Mimeo Available at SSRN:
https://ssrn.com/abstract=3582421

\item Kitahara, M., Okumura, Y. 2021. Improving Efficiency in School Choice
under Partial Priorities, International Journal of Game Theory 50, 971--987.

\item Kitahara, M, Okumura, Y. 2022. Stable Mechanisms in Controlled School
Choice. Mimeo Available at SSRN: https://ssrn.com/abstract=3806916

\item Kitahara, M., Okumura, Y. 2023. On Extensions of Partial Priorities in
School Choice. Mimeo Available at SSRN: https://ssrn.com/abstract=4462665

\item Kitahara, M., Okumura, Y. 2024. Extensions of Partial Priorities and
Stability in School Choice, Mathematical Social Sciences 131, 1-4.

\item Kojima, F. 2012. School choice: Impossibilities for affirmative
action. Games and Economic Behavior 75(2), 685--693.

\item Kwon, H., Shorrer, R. I. 2023. Justified-Envy-Minimal Efficient
Mechanisms for Priority-Based Matching, Mimeo

\item Kuvalekar, A. 2023. Matching with incomplete preferences, Mimeo
Available at https://arxiv.org/abs/2212.02613

\item Miyazaki, S., Okamoto, K. 2019. Jointly stable matchings. Journal of
Combinatorial Optimization 38(2), 646--665.

\item Ortega, J. 2018. Social integration in two-sided matching markets.
Journal of Mathematical Economics 78, 119--126.

\item Reny, P.J. 2022. Efficient Matching in the School Choice Problem.
American Economic Review 112(6), 2025--2043.

\item Roth, A.E. 1982. The economics of matching: stability and incentives,
Mathematics of Operations Research 7, 617-628.

\item S\"{o}nmez, T., Yenmez, M.B. 2022. Affirmative action in India via
vertical, horizontal, and overlapping reservations,\ Econometrica 90(3),
1143-1176.

\item Tang, Q., Yu, J., 2014. A new perspective on Kesten's school choice
with consent idea. Journal of Economic Theory 154, 543--561.

\item Tang, Q., Zhang, Y. 2021. Weak stability and Pareto efficiency in
school choice. Economic Theory 71, 533--552.

\item Troyan, P., Delacr\'{e}taz, D., Kloosterman, A. 2020. Essentially
stable matchings. Games and Economic Behavior 120, 370--390.

\item Wang, T. 2009. Preferential policies for minority college admission in
China: Recent developments, necessity, and impact,\ In M. Zhou and A.M. Hill
(eds) Affirmative action in China and the U.S., Palgrave Macmillan.\newpage
\end{description}

\section*{Appendices}

\subsection*{A.1. Weak M-stability}

Here, we introduce another fairness notion of the school choice problem with
multiple priority orders.

A matching $\mu $ is \textbf{weakly} \textbf{M-fair} for $\vartriangleright $
if for $\succ _{s}\in \mathcal{\vartriangleright }_{s}$, $\mu $ violates the
priority of $i$ for $s$ over $j$, then there is $\succ _{s}^{\prime }\in 
\mathcal{\vartriangleright }_{s}$ such that $\left( i,j\right) \notin \succ
_{s}^{\prime }$. In this notion, $\mu $ is regarded as unfair only if $%
sP_{i}\mu \left( i\right) \ $and $\left( i,j\right) \in \succ _{s}$ for 
\textit{all} $\succ _{s}\in \mathcal{\vartriangleright }_{s}$. A matching $%
\mu $ is said to be a \textbf{weakly\ M-stable matching }for $%
\vartriangleright $ if it is individually rational, non-wasteful and weakly
M-fair for $\vartriangleright $.

To consider the weakly\textbf{\ }M-stable matching, we let $w:2^{\mathcal{B}%
}\rightarrow \mathcal{B}$ be such that $\left( i,j\right) \in w\left( 
\mathcal{\vartriangleright }_{s}\right) $ if and only if $\left( i,j\right)
\in \succ _{s}$ for all $\succ _{s}\in \mathcal{\vartriangleright }_{s}$.
Trivially, $w\left( \mathcal{\vartriangleright }_{s}\right) $ is asymmetric
if $\succ _{s}\in \mathcal{B}$ for all $\succ _{s}\in \mathcal{%
\vartriangleright }_{s}$. Further, $W\left( \mathcal{\vartriangleright }%
\right) =\left( w\left( \mathcal{\vartriangleright }_{s}\right) \right)
_{s\in S}$.

\begin{proposition}
A matching $\mu $ is weakly M-stable for $\vartriangleright $ if and only if
it is stable for $W\left( \mathcal{\vartriangleright }\right) $.
\end{proposition}

\textbf{Proof. }First, suppose that $\mu $ is not weakly M-fair for $%
\vartriangleright $. Then, there is $(i,j,s)$ such that $\mu \left( j\right)
=s$, $sP_{i}\mu \left( i\right) $ and $\left( i,j\right) \in \succ _{s}$ for
all $\succ _{s}\in \mathcal{\vartriangleright }_{s}$. Then, $\left(
i,j\right) \in w\left( \mathcal{\vartriangleright }_{s}\right) $ and thus it
is not fair for $W\left( \mathcal{\vartriangleright }\right) $.

Second, suppose that $\mu $ is not stable for $W\left( \mathcal{%
\vartriangleright }\right) $. Then, there is $(i,j,s)$ such that $\mu \left(
j\right) =s$, $sP_{i}\mu \left( i\right) $ and $\left( i,j\right) \in
w\left( \mathcal{\vartriangleright }_{s}\right) $, which implies $\left(
i,j\right) \in \succ _{s}$ for all $\succ _{s}\in \mathcal{\vartriangleright 
}_{s}$. Therefore, $\mu $ is not weakly M-fair for $\vartriangleright $. 
\textbf{Q.E.D.}

\begin{proposition}
If $\succ _{s}$ is a partial order for all $\succ _{s}\in \mathcal{%
\vartriangleright }_{s}$, then $w\left( \mathcal{\vartriangleright }%
_{s}\right) $ is a partial order.
\end{proposition}

\textbf{Proof. }Suppose $(i,j)\in w\left( \mathcal{\vartriangleright }%
_{s}\right) $ and $(j,k)\in w\left( \mathcal{\vartriangleright }_{s}\right) $%
. Then, $(i,j)\in \succ _{s}$ and $(j,k)\in \succ _{s}$ for all $\succ
_{s}\in \mathcal{\vartriangleright }_{s}$. By the transitivity of $\succ
_{s} $, $(i,k)\in \succ _{s}$ for all $\succ _{s}\in \mathcal{%
\vartriangleright }_{s}\ $and thus $(i,j)\in w\left( \mathcal{%
\vartriangleright }_{s}\right) $. \textbf{Q.E.D.}

\begin{corollary}
If $\succ _{s}$ is a partial order for all $\succ _{s}\in \mathcal{%
\vartriangleright }_{s}$, then there is a weakly M-fair matching for $%
\mathcal{\vartriangleright }$.
\end{corollary}

Therefore, even if all elements in $\mathcal{\vartriangleright }_{s}$ are
partial orders for all $s\in S$, we can derive our results by using this
fairness notion.

\subsection*{A.2. Formal description of EADA algorithms}

To provide the proofs of Theorems 1 and 2, we firstly introduce a formal
description of the EADA algorithms that are briefly introduced in Section 5.

Let 
\begin{equation*}
\hat{G}_{S}=\left( \hat{I},\hat{S},\hat{P},\hat{\succ},\hat{q}\right) \in
2^{I}\times 2^{S}\times \mathbb{P}^{\left\vert I\right\vert }\times \mathcal{%
L}^{\left\vert S\right\vert }\times \mathbb{N}^{^{\left\vert S\right\vert }}
\end{equation*}%
be a problem where $\hat{\succ}_{s}$ is a linear order for all $s\in S$.
First, we introduce the Gale and Shapley's (1962) SPDA algorithm for $\hat{G}%
_{S}$.

\begin{description}
\item[Step $t$] Choose one student $i$ from $\hat{I}$ who is not tentatively
matched and has not been rejected by all of their acceptable schools in $%
\hat{S}$ at $\hat{P}_{i}$ yet. If there is no such student in $\hat{I}$,
then the algorithm terminates. The chosen student $i$ applies to the most
preferred school $s$ at $\hat{P}_{i}$ among $\hat{S}$ that have not rejected 
$i$ so far. If there are less than $\hat{q}_{s}$ students who are
tentatively matched to $s$, then $i$ is tentatively matched and go to the
next step. If otherwise; that is, if there are $\hat{q}_{s}$ students who
are tentatively matched to $s$, then the student with the lowest priority at 
$\hat{\succ}_{s}$ among the $\hat{q}_{s}$ students and $i$ is rejected and
the others are tentatively matched to $s$ and go to the next step.
\end{description}

Let $DA\left( \hat{G}_{S}\right) $ be the result of this algorithm for $\hat{%
G}_{S}$. A school $s\in \hat{S}$ is said to be \textbf{underdemanded} at a
matching $\mu $ with $\hat{G}_{S}$ if $\mu \left( i\right) \hat{R}_{i}s$ for
all $i\in \hat{I}$. Then, we have the following result.

\begin{remark}
A school $s$ is underdemanded at $DA\left( \hat{G}_{S}\right) $ with $\hat{G}%
_{S}$\ if and only if $s$ never rejects in any steps of the SPDA algorithm
for $\hat{G}_{S}$.
\end{remark}

Next, we introduce the EADA algorithms for $G_{S}=\left( I,S,P,\succ
,q\right) $.

\begin{description}
\item[Round $0$] Let $\succ ^{\ast }=\left( \succ _{s}^{\ast }\right) _{s\in
S}\in \mathcal{L}^{\left\vert S\right\vert }$ be such that $\succ _{s}^{\ast
}$ is an extension of $\succ _{s}$ for all $s\in S$.

\item[Round $1$] Let $I^{1}=I,$ $S^{1}=S$ and $P^{1}=P$. Run the SPDA for $%
G_{S}^{1}=\left( I^{1},S^{1},P^{1},\succ ^{\ast },q\right) $. Let $U^{1}$ be
the set of underdemanded schools at $DA\left( G_{S}^{1}\right) $; that is,
each $s\in U^{1}$ never rejects any student throughout the SPDA of this
Round. Moreover, let 
\begin{equation*}
E^{1}=\bigcup\nolimits_{s\in U^{1}\cup \left\{ \emptyset \right\} }DA\left(
G_{S}^{1}\right) \left( s\right) ,
\end{equation*}%
which is the set of students who is matched to an underdemanded school at $%
DA\left( G_{S}^{1}\right) $.

\item[Round $k$] Let $I^{k}=I^{k-1}\setminus E^{k-1}$ and $%
S^{k}=S^{k-1}\setminus U^{k-1}$. For all $i\in I^{k},$ let 
\begin{equation*}
Z_{i}^{k}=\left\{ s\in S^{k}\text{ }\left\vert \text{ }sP_{i}^{k-1}\emptyset
,\text{ }\left( j,i\right) \in \succ _{s}\text{ and }sP_{j}DA\left(
G_{S}^{k-1}\right) ,\text{ for some }j\in E^{k-1}\right. \right\} \text{,}
\end{equation*}%
and let $P_{i}^{k}$ be a linear order such that for all $s\in Z_{i}^{k}$, $%
\emptyset P_{i}^{k}s$ and for all $s^{\prime },s^{\prime \prime }\in \left(
S\cup \left\{ \emptyset \right\} \right) \setminus Z_{i}^{k},$ $s^{\prime
}P_{i}^{k-1}s^{\prime \prime }$ implies $s^{\prime }P_{i}^{k}s^{\prime
\prime }$. Run the SPDA for $G_{S}^{k}=\left( I^{k},S^{k},P^{k},\succ ^{\ast
},q\right) $. Let $U^{k}$ be the set of underdemanded schools at $DA\left(
G_{S}^{k}\right) $; that is, each $s\in U^{k}$ never rejects any student
throughout the SPDA of this Round. Moreover, let 
\begin{equation*}
E^{k}=\bigcup\nolimits_{s\in U^{k}\cup \left\{ \emptyset \right\} }DA\left(
G_{S}^{k}\right) \left( s\right) ,
\end{equation*}%
which is the set of students who is matched to an underdemanded school (or
unmatched) at $DA\left( G_{S}^{k}\right) $.\footnote{%
In each Round $k,$ $E^{k}$ is not empty. First, if no student is rejected,
then all schools are underdemanded and $E^{k}=I^{k}$. Otherwise, then we can
let $i$ be the student who is lastly rejected by a school. Then, $i$ must be
in $E^{k}$.}\newline
\end{description}

For notational convenience, let $\kappa :I\cup S\rightarrow \mathbb{N}$
satisfying $\kappa \left( i\right) =k$ for all $i\in E^{k}$ and $\kappa
\left( s\right) =k$ for all $s\in U^{k}$; that is, $\kappa \left( i\right) $
and $\kappa \left( s\right) $ represent the Rounds in which $i$ and $s$ are
eliminated, respectively. Moreover, let $\mathcal{E}^{k}\mathcal{=}%
\bigcup\nolimits_{k^{\prime }=1}^{k}E^{k^{\prime }},$ which is the set of
students eliminated from Round $1$ to Round $k$.

We let $\hat{\mu}^{k}$ be the matching obtained after Round $k$; that is, 
\begin{eqnarray*}
\hat{\mu}^{k}\left( i\right) &=&DA\left( G_{S}^{\kappa \left( i\right)
}\right) \text{ for all }i\in I\setminus I^{k}\text{ }(\kappa \left(
i\right) <k),\text{ } \\
\hat{\mu}^{k}\left( i\right) &=&DA\left( G_{S}^{k}\right) \text{ for all }%
i\in I^{k},\text{ }(\kappa \left( i\right) \geq k).
\end{eqnarray*}%
That is, for each student who is remaining in Round $k$, $\hat{\mu}^{k}$ is
the SPDA result in Round $k$, and for each student who is not remaining, it
is the SPDA result in the round when the student is eliminated. In the
terminal step $K$, $S=\bigcup\nolimits_{k=1}^{K}U^{k}$ and the resulting
matching $\hat{\mu}^{K}$.

\subsection*{A.3. Proof of Theorem 1.}

For $\succ \in \mathcal{P}^{n}$, we fix $\succ _{s}^{\ast }$ as an arbitrary
extension of $\succ _{s}$ for all $s\in S$. For notational simplicity, let $%
EA\left( \succ ,\succ ^{\ast }\right) =\mu ^{\ast }$. First, we show the
following three Claims.

\begin{claim}
If $\hat{\mu}^{k-1}$ is stable for $\succ $, then$\ \hat{\mu}^{k}\left(
i\right) R_{i}\hat{\mu}^{k-1}\left( i\right) $ for all $i\in I$.
\end{claim}

\textbf{Proof.} Suppose that $\hat{\mu}^{k-1}$ is stable for $\succ $. If no
student is rejected by any school in the SPDA of Round $k-1$, then $\hat{\mu}%
^{k}\left( i\right) =\hat{\mu}^{k-1}\left( i\right) $ for all $i\in I$ and
the proof is finished. Therefore, we assume that some student is rejected by
some school in the SPDA of Round $k-1$.

Suppose not; that is, 
\begin{equation*}
I^{\prime }=\left\{ i\in I\text{ }\left\vert \text{ }\hat{\mu}^{k-1}\left(
i\right) P_{i}\hat{\mu}^{k}\left( i\right) \right. \right\} \neq \emptyset .
\end{equation*}%
Since $\hat{\mu}^{k}\left( i\right) =\hat{\mu}^{k-1}\left( i\right) $ for
all $i\in \mathcal{E}^{k-1}$, $I^{\prime }\subseteq I\setminus \mathcal{E}%
^{k-1}$; that is, any student in $I^{\prime }$ has not eliminated yet. Let $%
i\in I^{\prime }$ be the student who is rejected by $\hat{\mu}^{k-1}\left(
i\right) =DA\left( G_{S}^{k-1}\right) \left( i\right) $ in the \textit{%
earliest step} of the SPDA in Round $k$ among $I^{\prime }$. Let $s=DA\left(
G_{S}^{k-1}\right) \left( i\right) $ and $t$ be that step. Then, in Step $t$
(of the SPDA in Round $k$), there are $q_{s}$ students who is temporarily
accepted by $s$. Moreover, for any student $j$ who is temporarily accepted
by $s$ in Step $t$, $sR_{j}DA\left( G_{S}^{k}\right) \left( j\right) $ and $%
\left( j,i\right) \in \succ _{s}^{\ast }$. On the other hand, since $i\in 
\hat{\mu}^{k-1}\left( s\right) $, there is $j\notin \hat{\mu}^{k-1}\left(
s\right) $ who is the temporarily accepted students in Step $t$ of the SPDA
in Round $k$. Since $\hat{\mu}^{k-1}$ is stable, $DA\left(
G_{S}^{k-1}\right) \left( j\right) P_{j}s$ and therefore $j\in I^{\prime }$.
Since $j$ is temporarily accepted by $s$ in Step $t$, $j$ is rejected by $%
DA\left( G_{S}^{k-1}\right) \left( j\right) $ before Step $t$. Since $i$ is
the student who is firstly rejected by $DA\left( G_{S}^{k-1}\right) \left(
i\right) $, this is a contradiction. \textbf{Q.E.D. }\newline

\begin{claim}
If $\hat{\mu}^{k-1}$ is stable for $\succ $, then $\left\vert \hat{\mu}%
^{k-1}\left( s\right) \right\vert =\left\vert \hat{\mu}^{k}\left( s\right)
\right\vert $ for all $s\in S$.
\end{claim}

\textbf{Proof.} Suppose that $\hat{\mu}^{k-1}$ is stable for $\succ $. By
Claim 1 and the individually rationality of $\hat{\mu}^{k-1}$, for any $i$
such that $\hat{\mu}^{k-1}\left( i\right) \in S$, we have $\hat{\mu}%
^{k}\left( i\right) R_{i}\hat{\mu}^{k-1}\left( i\right) P_{i}\emptyset $ and
thus $\hat{\mu}^{k}\left( i\right) \in S$. Therefore, 
\begin{equation}
\sum_{s\in S}\left\vert \hat{\mu}^{k-1}\left( s\right) \right\vert \leq
\sum_{s\in S}\left\vert \hat{\mu}^{k}\left( s\right) \right\vert .  \label{b}
\end{equation}

Now, suppose not; that is, $\left\vert \hat{\mu}^{k-1}\left( s\right)
\right\vert \neq \left\vert \hat{\mu}^{k}\left( s\right) \right\vert $ for
some $s\in S$. First, we assume $\left\vert \hat{\mu}^{k-1}\left( s\right)
\right\vert <\left\vert \hat{\mu}^{k}\left( s\right) \right\vert \leq q_{s}$
for some $s\in S$. Then, there is $i\in \hat{\mu}^{k}\left( s\right)
\setminus \hat{\mu}^{k-1}\left( s\right) $. By Claim 1, $sP_{i}\hat{\mu}%
^{k-1}\left( i\right) $. However, $\left\vert \hat{\mu}^{k-1}\left( s\right)
\right\vert <q_{s}$ and $sP_{i}\hat{\mu}^{k-1}\left( i\right) $ contradict
the nonwastefulness of $\hat{\mu}^{k-1}$. Therefore, $\left\vert \hat{\mu}%
^{k-1}\left( s\right) \right\vert \geq \left\vert \hat{\mu}^{k}\left(
s\right) \right\vert $ for all $s\in S$ and $\left\vert \hat{\mu}%
^{k-1}\left( s\right) \right\vert >\left\vert \hat{\mu}^{k}\left( s\right)
\right\vert $ for some $s\in S$. However, these contradict (\ref{b}). 
\textbf{Q.E.D.}

\begin{claim}
If $\hat{\mu}^{k-1}$ is stable for $\succ $, then$\ \hat{\mu}^{k}$ is also
stable for $\succ $.
\end{claim}

\textbf{Proof.} Suppose that\textbf{\ }$\hat{\mu}^{k-1}$ is stable for $%
\succ $. First, since $\hat{\mu}^{k-1}$ is individually rational, by Claim
1, $\hat{\mu}^{k}$ is also individual rational.

Second, we show the nonwastefulness of $\hat{\mu}^{k}$. By Claim 1, $sP_{i}%
\hat{\mu}^{k}\left( i\right) $ implies $sP_{i}\hat{\mu}^{k-1}\left( i\right) 
$. Moreover, since $\hat{\mu}^{k-1}$ is nonwasteful, $sP_{i}\hat{\mu}%
^{k-1}\left( i\right) $ also implies $\left\vert \hat{\mu}^{k-1}\left(
s\right) \right\vert =q_{s}$. Thus, $sP_{i}\hat{\mu}^{k}\left( i\right) $
implies $\left\vert \hat{\mu}^{k-1}\left( s\right) \right\vert =q_{s}$.
However, by Claim 2, $\left\vert \hat{\mu}^{k-1}\left( s\right) \right\vert
=\left\vert \hat{\mu}^{k}\left( s\right) \right\vert =q_{s}$ and thus $\hat{%
\mu}^{k}$ is also nonwasteful.

Third, we show the fairness of $\hat{\mu}^{k}$. Suppose not; that is, we can
let $k$ be the smallest integer where $\hat{\mu}^{k-1}$ is stable for $\succ
,$ but $\hat{\mu}^{k}$ is not fair; that is, there are $i$ and $j$ such that 
$\hat{\mu}^{k}\left( j\right) =sP_{i}\hat{\mu}^{k}\left( i\right) \ $and $%
\left( i,j\right) \in \succ _{s}$. Since $\hat{\mu}^{1}=\mu ^{1}$ must be
stable for $\succ $, $\hat{\mu}^{1},\hat{\mu}^{2},\cdots ,\hat{\mu}^{k-1}$
are stable for $\succ $.

First, we show that $i\in I^{k}$; that is, $i$ has not eliminated at Round $%
k $ yet. Suppose not; that is, $i\in \mathcal{E}^{k-1}$. If $j\in \mathcal{E}%
^{k-1}$; that is, if $j\,$and $s\left( =\hat{\mu}^{k}\left( j\right) \right) 
$ have already eliminated at Round $k$, then $\hat{\mu}^{k}\left( j\right) =%
\hat{\mu}^{k-1}\left( j\right) =s$ and $\hat{\mu}^{k}\left( i\right) =\hat{%
\mu}^{k-1}\left( i\right) $. However, these contradict the stability of $%
\hat{\mu}^{k-1}$. Hence $j\in I^{k}\,$and $s$ have not eliminated at Round $%
k $ yet. Since $\left( i,j\right) \in \succ _{s}$, by the description of the
EADA algorithm above, $\emptyset P_{j}^{k}s$, contradicting $\hat{\mu}%
^{k}\left( j\right) =s$. Thus, $i\in I^{k}$.

Second, suppose $sP_{i}^{k}\emptyset $. Then, since $\hat{\mu}^{1},\hat{\mu}%
^{2},\cdots ,\hat{\mu}^{k-1}$ are stable for $\succ $, and Claim 1 holds, $%
sP_{i}\hat{\mu}^{k}\left( i\right) R_{i}\hat{\mu}^{k-1}\left( i\right)
R_{i}\cdots R_{i}\hat{\mu}^{1}\left( i\right) $. If $s$ is eliminated at an
earlier Round $\kappa \left( s\right) <k$, then it is underdemanded in the
Round $\kappa \left( s\right) $ which implies $\hat{\mu}^{\kappa \left(
s\right) }\left( i\right) R_{i}s$ contradicting $sP_{i}\hat{\mu}^{\kappa
\left( s\right) }\left( i\right) $. Therefore, $s$ and $i$ have not
eliminated at Round $k$ yet. Then, both $i$ and $j$ apply to $s$ in the SPDA
of Round $k$. Thus, $\left( j,i\right) \in \succ _{s}^{\ast }$. However,
since $\succ _{s}^{\ast }$ is an extension of $\succ _{s}$, $\left(
j,i\right) \in \succ _{s}^{\ast }$ contradicts $\left( i,j\right) \in \succ
_{s}$.

Therefore, $\emptyset P_{i}^{k}s$. Since $sP_{i}\emptyset $, there is $%
i^{\prime }\in \mathcal{E}^{k-1}$ such that $sP_{i^{\prime }}\hat{\mu}%
^{k}\left( i^{\prime }\right) =\hat{\mu}^{k-1}\left( i^{\prime }\right) $
and $\left( i^{\prime },i\right) \in \succ _{s}$. By the transitivity of $%
\succ _{s}$, $\left( i^{\prime },i\right) \in \succ _{s}$ and $\left(
i,j\right) \in \succ _{s}$ imply $\left( i^{\prime },j\right) \in \succ _{s}$%
.\footnote{%
When $\succ _{s}$ is not transitive for some $s\in S$, \ $\hat{\mu}^{k}$ may
not be fair for $\succ $ even if $\hat{\mu}^{k-1}$ is fair for $\succ $. In
Appendix A.4., we introduce an example that $\succ _{s}$ is not transitive
for some $s\in S$ and $\hat{\mu}^{k}$ is not fair for $\succ $ even if $\hat{%
\mu}^{k-1}$ is fair for $\succ $.\ } However, by the first part of this
proof, $i^{\prime }\in \mathcal{E}^{k-1}$, $\left( i^{\prime },j\right) \in
\succ _{s}$ and $\hat{\mu}^{k}\left( j\right) =sP_{i^{\prime }}\hat{\mu}%
^{k}\left( i^{\prime }\right) $ are not compatible. Thus, we have Claim 3. 
\textbf{Q.E.D. }\newline

Since $\hat{\mu}^{1}=\mu ^{1}$ is stable for $\succ $ and Claim 3 is
satisfied, $\hat{\mu}^{1},\cdots ,\hat{\mu}^{K}\left( =\mu ^{\ast }\right) $
are stable. By Claim 1, 
\begin{equation*}
\hat{\mu}^{K}\left( i\right) R_{i}\hat{\mu}^{K-1}\left( i\right) R_{i}\hat{%
\mu}^{k-2}\left( i\right) \cdots R_{i}\hat{\mu}^{1}\left( i\right) \text{.}
\end{equation*}%
Moreover, by Claim 2, 
\begin{equation*}
\left\vert \hat{\mu}^{1}\left( s\right) \right\vert =\cdots =\left\vert \hat{%
\mu}^{K}\left( s\right) \right\vert ,
\end{equation*}%
for all $s\in S$.

Now, we show Theorem 1. Suppose not; that is, there is a stable matching $%
\nu $ for $\succ $ that Pareto dominates $\mu ^{\ast }\left( =\hat{\mu}%
^{K}\right) $. Let 
\begin{equation*}
i\in I^{\ast }=\left\{ i\in I\text{ }\left\vert \text{ }\nu \left( i\right)
P_{i}\mu ^{\ast }\left( i\right) \right. \right\} \neq \emptyset .
\end{equation*}%
First, show that there exist $i^{\prime }\in I^{\ast }$ such that $\kappa
\left( \nu \left( i^{\prime }\right) \right) >\kappa \left( i^{\prime
}\right) $; that is, the step which $\nu \left( i^{\prime }\right) $ is not
eliminated is after that which $i^{\prime }$ is not eliminated. Let $i\in
I^{\ast }$ and $\nu \left( i\right) =s$. Here, we assume $\kappa \left(
s\right) \leq \kappa \left( i\right) $. In this case, since $s$ is
underdemanded at $\mu ^{\kappa \left( s\right) },$ $i$ does not apply to $s$
in the SPDA in Round $\kappa \left( s\right) $, although $i$ has not
eliminated yet. Moreover, by Claim 1, $sP_{i}\mu ^{\ast }\left( i\right)
R_{i}\hat{\mu}^{\kappa \left( s\right) }$. Thus, $\emptyset P_{i}^{\kappa
\left( s\right) }s$ must be satisfied; that is, there is $i^{\prime }$ such
that $\kappa \left( i^{\prime }\right) <\kappa \left( s\right) $, $%
sP_{i^{\prime }}\mu ^{\ast }\left( i^{\prime }\right) $ and $\left(
i^{\prime },i\right) \in \succ _{s}$. Since $\nu $ is stable for $\succ $
and $\nu \left( i\right) =s$, $\nu \left( i^{\prime }\right) R_{i^{\prime
}}sP_{i^{\prime }}\mu ^{\ast }\left( i^{\prime }\right) $ and thus $%
i^{\prime }\in I^{\ast }$. Therefore, for any $i\in I^{\ast }$ such that $%
\kappa \left( \nu \left( i\right) \right) (=\kappa \left( s\right) )\leq
\kappa \left( i\right) ,$ there is another $i^{\prime }\in I^{\ast }$ such
that $\kappa \left( i^{\prime }\right) <\kappa \left( i\right) $. Since $%
I^{\ast }$ is finite, there must exist $i^{\prime }\in I^{\ast }$ such that $%
\kappa \left( \nu \left( i^{\prime }\right) \right) >\kappa \left( i^{\prime
}\right) $.

Let $i^{\prime }\in I^{\ast }$ be such that $\kappa \left( i^{\prime
}\right) \leq \kappa \left( i\right) $ for all $i\in I^{\ast }$. Then $%
\kappa \left( \nu \left( i^{\prime }\right) \right) >\kappa \left( i^{\prime
}\right) $, because of the fact above. Let $\kappa \left( i^{\prime }\right)
=k$. By Claims 1 and 3, $\hat{\mu}^{k}$ is stable and Pareto dominated by $%
\nu \left( i^{\prime }\right) $. Moreover, by Claim 1, 
\begin{equation*}
I^{\ast }\subseteq \left\{ i\in I\text{ }\left\vert \text{ }\nu \left(
i\right) P_{i}\hat{\mu}^{k}\left( i\right) \right. \right\} =I^{\prime }
\end{equation*}%
and thus, $i^{\prime }\in I^{\prime }$.

We consider two subsets of students; 
\begin{eqnarray*}
I_{1} &=&\left\{ i\in I\text{ }\left\vert \text{ }\nu \left( i\right) \in
S^{k+1}\text{ and }\hat{\mu}^{k}\left( i\right) \in \left( S\cup \left\{
\emptyset \right\} \right) \setminus S^{k+1}\right. \right\} , \\
I_{2} &=&\left\{ i\in I\text{ }\left\vert \text{ }\nu \left( i\right) \in
\left( S\cup \left\{ \emptyset \right\} \right) \setminus S^{k+1}\text{ and }%
\hat{\mu}^{k}\left( i\right) \in S^{k+1}\right. \right\} .
\end{eqnarray*}%
Then, since $\kappa \left( \nu \left( i^{\prime }\right) \right) >\kappa
\left( i^{\prime }\right) $, $\nu \left( i^{\prime }\right) \in S^{k+1}$ and 
$i^{\prime }$ is eliminated in Step $k$, $i^{\prime }\in I_{1}$ and hence $%
I_{1}$ is not empty. We show that $I_{2}$ is empty. Suppose not; that is, $%
i^{\prime \prime }\in I_{2}$. Since $\nu $ Pareto dominates $\hat{\mu}^{k}$
and $\hat{\mu}^{k}$ is individually rational, $\nu \left( i^{\prime \prime
}\right) \in S$. Then, $\nu \left( i^{\prime \prime }\right) $ has already
eliminated but $i^{\prime \prime }$ has not eliminated yet at $k$. Let $%
\kappa \left( \nu \left( i^{\prime \prime }\right) \right) =k^{\prime \prime
}\left( <k\right) .$ Since $\nu \left( i^{\prime \prime }\right) $ is
underdemanded in the SPDA of Round $k^{\prime \prime }$, $i^{\prime \prime }$
does not apply to $\nu \left( i^{\prime \prime }\right) $. Then, since $\nu
\left( i^{\prime \prime }\right) P_{i^{\prime \prime }}\emptyset $, we have $%
\emptyset P_{i^{\prime \prime }}^{k^{\prime \prime }}\nu \left( i^{\prime
\prime }\right) $. Thus, there is $\kappa \left( i^{\prime \prime \prime
}\right) <k^{\prime },$ $\nu \left( i^{\prime \prime }\right) P_{i^{\prime
\prime \prime }}\hat{\mu}^{k}\left( i^{\prime \prime \prime }\right) $ and $%
\left( i^{\prime \prime \prime },i^{\prime \prime }\right) \in \succ _{\nu
\left( i^{\prime \prime }\right) }$. Since $\nu $ is fair for $\succ $, $\nu
\left( i^{\prime \prime \prime }\right) R_{i^{\prime \prime \prime }}\nu
\left( i^{\prime \prime }\right) $ and thus $i^{\prime \prime \prime }\in
I^{\prime }$. Moreover, since $\kappa \left( i^{\prime \prime \prime
}\right) <k^{\prime \prime }$, $\hat{\mu}^{k}\left( i^{\prime \prime \prime
}\right) =\hat{\mu}^{\ast }\left( i^{\prime \prime \prime }\right) $ and
thus $i^{\prime \prime \prime }\in I^{\ast }$. However, this contradicts $%
\kappa \left( i^{\prime }\right) \leq \kappa \left( i\right) $ for all $i\in
I^{\ast }$. Therefore, $I_{2}=\emptyset $.

Then, since $\left\vert I_{1}\right\vert \geq 1$, there is at least one
student $i$ such that $\nu \left( i\right) \in S^{k+1}$ and $\hat{\mu}%
^{k}\left( i\right) \notin S^{k+1}$. However, since $\left\vert
I_{2}\right\vert =0$, there is no student $i$ such that $\nu \left( i\right)
\notin S^{k+1}$ and $\hat{\mu}^{k}\left( i\right) \in S^{k+1}$. Thus, there
is $s\in S^{k+1}$ such that $\left\vert \hat{\mu}^{k}\left( s\right)
\right\vert <\left\vert \nu \left( s\right) \right\vert \leq q_{s}$ and thus
there is $i^{\prime }\in \nu \left( s\right) \setminus \hat{\mu}^{k}\left(
s\right) $. Since $\nu $ Pareto dominates $\hat{\mu}^{k}$, $s=\nu \left(
i^{\prime }\right) P_{i^{\prime }}\hat{\mu}^{k}\left( i^{\prime }\right) ,$
which contradicts the nonwastefulness of $\hat{\mu}^{k}$. \textbf{Q.E.D.}

\subsection*{A.4. Intransitive example}

By Theorem 1, for any $\succ \in \mathcal{P}^{n}$, $EA\left( \succ ,\succ
^{\ast }\right) $ is an SOSM for $\succ $. On the other hand, if $\succ
_{s}\notin \mathcal{P}$ for some $s$, $EA\left( \succ ,\succ ^{\ast }\right) 
$ may not be stable for $\succ $. This result holds even if $\succ _{s}$ is
acyclic for all $s\in S$. Here, we provide an example to show this fact.

Let $I=\left\{ i_{1},\cdots ,i_{5}\right\} $ and $S=\left\{ s_{1},\cdots
,s_{4}\right\} $, $q_{s}=1$ for all $s\in S,$ and 
\begin{gather*}
P_{i_{1}}:s_{2}\text{ }s_{1}\text{ }\emptyset ,\,P_{i_{2}}:s_{1}\text{ }s_{2}%
\text{ }\emptyset ,\,P_{i_{3}}:s_{1}\text{ }s_{4}\text{ }s_{3}\text{ }%
\emptyset ,\text{ } \\
P_{i_{4}}:s_{3}\text{ }s_{4}\text{ }\emptyset ,\,P_{i_{5}}:s_{1}\text{ }s_{4}%
\text{ }\emptyset , \\
\succ _{s_{1}}=\left\{ \left( i_{1},i_{5}\right) ,\left( i_{1},i_{3}\right)
,\left( i_{5},i_{3}\right) ,\left( i_{3},i_{2}\right) \right\} ,\text{ }%
\succ _{s_{2}}:\text{ }i_{2}\text{ }i_{1}\text{ }i_{3}\text{ }i_{4}\text{ }%
i_{5} \\
\succ _{s_{3}}:i_{3}\text{ }i_{4}\text{ }i_{2}\text{ }i_{1}\text{ }i_{5},%
\text{ }\succ _{s_{4}}:i_{4}\text{ }\left[ i_{3}\text{ }i_{5}\right] \text{ }%
i_{1}\text{ }i_{2}.
\end{gather*}%
Since $\left( i_{5},i_{3}\right) ,\left( i_{3},i_{2}\right) \in \succ
_{s_{1}}$ but $\left( i_{5},i_{2}\right) \notin \succ _{s_{1}}$, $\succ
_{s_{1}}$ is not transitive and thus not a partial order. However, $\succ
_{s_{1}}$ is acyclic, because $\left( i_{2},i_{5}\right) \notin \succ
_{s_{1}}$.

Let 
\begin{gather*}
\succ _{s_{1}}^{\ast }:i_{1}\text{ }i_{5}\text{ }i_{3}\text{ }i_{2}\text{ }%
i_{4},\text{ }\succ _{s_{2}}^{\ast }=\succ _{s_{2}} \\
\succ _{s_{3}}^{\ast }=\succ _{s_{3}}\text{ }\succ _{s_{4}}^{\ast }:i_{4}%
\text{ }i_{5}\text{ }i_{3}\text{ }i_{1}\text{ }i_{2}.
\end{gather*}%
We derive $EA\left( \succ ,\succ ^{\ast }\right) $. In Round 1, 
\begin{equation*}
DA\left( G_{S}^{1}\right) \left( i_{l}\right) =s_{l},\text{ for }l=1,2,3,4,%
\text{ }DA\left( G_{S}^{1}\right) \left( i_{5}\right) =\emptyset .
\end{equation*}%
Then, $E^{1}=\left\{ i_{5}\right\} $; that is, $i_{5}$ is eliminated and
their assignment is $\emptyset $. Then, since $\left( i_{5},i_{3}\right) \in
\succ _{s_{1}}$ and $\left( i_{5},i_{3}\right) \in \succ _{s_{4}}$, $%
Z_{i_{3}}^{2}=\left\{ s_{1},s_{4}\right\} $. Hence $s_{3}P_{i_{3}}^{2}%
\emptyset P_{i_{3}}^{2}s_{l}$ for all $l=1,2,4$, to prevent $i_{3}$ being
matched to $s_{1}$ or $s_{4}$.

Second, 
\begin{equation*}
DA\left( G_{S}^{2}\right) \left( i_{1}\right) =s_{2},\text{ }DA\left(
G_{S}^{2}\right) \left( i_{2}\right) =s_{1},\text{ }DA\left(
G_{S}^{2}\right) \left( i_{3}\right) =s_{3},\text{ }DA\left(
G_{S}^{2}\right) \left( i_{4}\right) =s_{4}.
\end{equation*}%
Then, $E^{2}=\left\{ i_{1},i_{2},i_{4}\right\} $ and $U^{2}=\left\{
s_{1},s_{2},s_{4}\right\} $.

Finally, $DA\left( G_{S}^{3}\right) \left( i_{3}\right) =s_{3}$. Therefore, $%
EA\left( \succ ,\succ ^{\ast }\right) =\mu ^{\ast }$ is such that%
\begin{equation*}
\left( EA\left( \succ ,\succ ^{\ast }\right) \left( i_{1}\right) ,\cdots
,EA\left( \succ ,\succ ^{\ast }\right) \left( i_{5}\right) \right) =\left(
s_{2},s_{1},s_{3},s_{4},\emptyset \right) \text{.}
\end{equation*}%
However, $EA\left( \succ ,\succ ^{\ast }\right) $ is not fair for $\succ ,$
because $\left( i_{3},i_{2}\right) \in \succ _{s_{1}}$ and $%
s_{1}P_{i_{3}}s_{3}$.

To understand why $EA\left( \succ ,\succ ^{\ast }\right) $ fails to be fair
for $\succ $. Let $\succ _{s_{1}}^{\prime }=\succ _{s_{1}}\cup \left\{
\left( i_{5},i_{2}\right) \right\} $ and $\succ _{s_{l}}^{\prime }=\succ
_{s_{l}}$ for all $l=2,\ldots ,5$. Then, $\succ _{s_{1}}^{\prime }$ a
partial order. Suppose that $\succ _{s_{1}}^{\prime }$ is used instead of $%
\succ _{s_{1}}$. Then, $DA\left( G_{S}^{1}\right) $ for $\succ ^{\prime }$
is completely the same as that for $\succ $. However, since $\left(
i_{5},i_{2}\right) \in \succ _{s_{1}}$, $Z_{i_{2}}^{2}=\left\{ s_{1}\right\} 
$ and $\emptyset P_{i_{2}}^{2}s_{1}$. Therefore, $i_{2}$ is not matched to $%
s_{1}$ and\ we have 
\begin{equation*}
\left( EA\left( \succ ^{\prime },\succ ^{\ast }\right) \left( i_{1}\right)
,\cdots ,EA\left( \succ ^{\prime },\succ ^{\ast }\right) \left( i_{5}\right)
\right) =\left( s_{1},s_{2},s_{3},s_{4},\emptyset \right) \text{.}
\end{equation*}%
On the other hand, if $\succ $ is used, then $Z_{i_{2}}^{2}=\emptyset $
because $\left( i_{5},i_{2}\right) \notin \succ _{s_{1}}$. Therefore, in
that case, $i_{2}$ is matched to $s_{1}$ and $EA\left( \succ ,\succ ^{\ast
}\right) $ is not fair for $\succ $.

Hence, ensuring the transitivity of $\succ _{s}$ for each $s\in S$ is
important\ for $EA\left( \succ ,\succ ^{\ast }\right) $ to be fair for $%
\succ $.

\subsection*{A.5. Other definitions of improvements and responsiveness}

Here we consider two other notions of responsiveness, which are weaker and
stronger than the responsiveness discussed above.

First, we show that the notion of responsiveness introduced in Jiao and Shen
(2021) and Jiao et al. (2022). Jiao and Tian (2019) (as well as Jiao and
Shen (2021) and Jiao et al. (2022)) define a stronger notion of improvement
than ours. Thus, here, we say that $\succ ^{\prime }$ is a \textbf{strict} 
\textbf{improvement }of\textbf{\ }$\succ $ for $I^{\prime }\subseteq I$ if
for all $s\in S$, (S1) $\left( i,j\right) \in \succ _{s}$ and $i\in
I^{\prime }$ imply $\left( i,j\right) \in \succ _{s}^{\prime }$, and (S2) $%
\left( i,j\right) \in \succ _{s}$ and $i,j\in I\setminus I^{\prime }$ imply $%
\left( i,j\right) \in \succ _{s}^{\prime }$.\footnote{%
The notion of strict improvement introduced here is slightly more general
than that introduced by Jiao and Tian (2019).}

\begin{remark}
If $\succ ^{\prime }$ is a strict improvement\textbf{\ }of\textbf{\ }$\succ $
for $I^{\prime }\subseteq I$, then $\succ ^{\prime }$ is an improvement%
\textbf{\ }of\textbf{\ }$\succ $ for $I^{\prime }\subseteq I$.
\end{remark}

\textbf{Proof.} First, we show that if $\succ ^{\prime }$ is a strict
improvement\textbf{\ }of\textbf{\ }$\succ $ for $I^{\prime }\subseteq I$,
then $\succ ^{\prime }$ is an improvement\textbf{\ }of\textbf{\ }$\succ $
for $I^{\prime }\subseteq I$. Suppose that $\succ ^{\prime }$ is a strict
improvement\textbf{\ }of\textbf{\ }$\succ $ for $I^{\prime }\subseteq I$ and
there is some $j\in I$ and some $s\in S$ such that $\left( j,i\right) \in
\succ _{s}$ and $\left( i,j\right) \in \succ _{s}^{\prime }$. Since $\left(
i,j\right) \in \succ _{s}^{\prime }$ and $\succ _{s}^{\prime }\in \mathcal{L}
$, $\left( j,i\right) \notin \succ _{s}^{\prime }$. Moreover, $\left(
j,i\right) \in \succ _{s}$ and $\succ _{s}\in \mathcal{L}$, $\left(
i,j\right) \notin \succ _{s}$. By the contraposition of (S1), $i\in
I^{\prime }$. \textbf{Q.E.D.}\newline

We consider the difference between the definition of Jiao and Tian (2019)
and ours. Jiao and Tian (2019) require (S1) and (S2); that is, the
priorities among the minority students as well as those among the majority
students are not changed after the improvements. On the other hand, we allow
that the priorities among the majority students are changed after the
improvements.

To clarify the fact, we revisit Example 3. Although $\succ _{s}^{2}$ and $%
\succ _{s}^{4}$ are improvements of $\succ _{s}^{1}$ for the minority
students, neither $\succ _{s}^{2}$ nor $\succ _{s}^{4}$ is a strict
improvement\textbf{\ }of $\succ _{s}^{1}$ for the minority students. That
is, if the bonus point of a minority student is different from that of
another minority student, then $\succ ^{\prime }$ may not be a strict
improvement over $\succ $ for the minority students. On the other hand, in
our definition, bonus points are allowed to differ among minority students.%
\newline

Next, we have the following result.

\begin{remark}
Let $\succ ,\succ ^{\prime },\succ ^{\prime \prime }\in \mathcal{L}%
^{\left\vert S\right\vert }$ be such that $\succ ^{\prime }$ is a strict
improvement of $\succ $ and $\succ ^{\prime \prime }$\textbf{\ }for $%
I^{\prime }\subseteq I,$ and $\succ ^{\prime \prime }$ is a strict
improvement of $\succ $ for $I^{\prime }$. Then, $\succ ^{\prime }$ more
improves $\succ $ than $\succ ^{\prime \prime }$ for $I^{\prime }$.
\end{remark}

\textbf{Proof.} By Remark 6, we have that $\succ ^{\prime }$ is an
improvement of $\succ $ and $\succ ^{\prime \prime }$\textbf{\ }for $%
I^{\prime }\subseteq I,$ and $\succ ^{\prime \prime }$ is an improvement of $%
\succ $ for $I^{\prime }$; that is, (i) and (ii) are satisfied. We show that
(iii) is satisfied. Suppose $\left( i,i^{\prime }\right) \in \succ _{s}$ and 
$\left( i^{\prime },i\right) \in \succ _{s}^{\prime \prime }$. We show $%
\left( i^{\prime },i\right) \in \succ _{s}^{\prime }$. Since $\succ ^{\prime
\prime }$ is a strict improvement\textbf{\ }of\textbf{\ }$\succ ,$ $i\notin
I^{\prime }$ and $i^{\prime }\in I^{\prime }$. Since $\succ ^{\prime }$ is a
strict improvement of $\succ ^{\prime \prime }$ for $I^{\prime }$, $\left(
i^{\prime },i\right) \in \succ _{s}^{\prime \prime }$ and $i^{\prime }\in
I^{\prime }\,$imply $\left( i^{\prime },i\right) \in \succ _{s}^{\prime }$.
Thus, $\succ ^{\prime }$ more improves $\succ $ than $\succ ^{\prime \prime
} $ for $I^{\prime }$. \textbf{Q.E.D.}\newline

By Theorem 3 and Remark 6 we immediately have the following result.

\begin{corollary}
Let $\succ ,\succ ^{\prime },\succ ^{\prime \prime }\in \mathcal{L}%
^{\left\vert S\right\vert }$ be such that $\succ ^{\prime }$ and $\succ
^{\prime \prime }$ are strict improvements of $\succ $ for $I^{\prime
}\subseteq I$. If $\succ ^{\prime }$ is a strict improvement\textbf{\ }of%
\textbf{\ }$\succ ^{\prime \prime }$ for $I^{\prime }$, then $\phi ^{\ast
}\left( \succ ,\succ ^{\prime \prime }\right) $ is not Pareto dominated by $%
\phi ^{\ast }\left( \succ ,\succ ^{\prime }\right) $ for $I^{\prime }$.
\end{corollary}

In Jiao and Shen (2021) and Jiao et al. (2022), a mechanism is said to be 
\textbf{responsive }if the fact that $\succ ^{\prime \prime }$ is strict
improvement of $\succ ^{\prime }$ for $I^{\prime }$ implies that $\phi
^{\ast }\left( \succ ,\succ ^{\prime \prime }\right) $ is not Pareto
dominated by $\phi ^{\ast }\left( \succ ,\succ ^{\prime }\right) $ for $%
I^{\prime }$. By this result, $\phi ^{\ast }$ is responsive in the sense of
Jiao and Shen (2021) and Jiao et al. (2022). We revisit Example 3. The
responsiveness in the sense of Jiao and Shen (2021) and Jiao et al. (2022)
does not require that $\phi ^{\ast }\left( \succ ^{1},\succ ^{4}\right) $ is
not Pareto dominated by $\phi ^{\ast }\left( \succ ^{1},\succ ^{2}\right) $
for $\left\{ 1,2,3,4\right\} $. On the other hand, our responsiveness
requires it.

Next, we try to strengthen the responsiveness of improvements property. A
mechanism $\phi $ is said to be \textbf{strictly} \textbf{responsive to
improvements} if the facts $\succ ^{\prime }$ and $\succ ^{\prime \prime }$
are improvements of $\succ $ for $I^{\prime }\subseteq I,$ and $\succ
^{\prime }$ is an improvement of $\succ ^{\prime \prime }$ for $I^{\prime
}\subseteq I$ imply that $\phi \left( \succ ,\succ ^{\prime }\right) $ is
not Pareto dominated for $I^{\prime }$ by $\phi \left( \succ ,\succ ^{\prime
\prime }\right) $. Note that if $\succ ,$ $\succ ^{\prime }$ and $\succ
^{\prime \prime }$ satisfy the facts in the definition, $\succ ^{\prime }$
may not more improve $\succ $ than $\succ ^{\prime \prime }$. This is
because, (iii) in the definition of \textquotedblleft more
improvement\textquotedblright\ is not satisfied.

We show that the strict responsiveness is not compatible with the student
optimal M-stability. Let $I=\left\{ i_{1},i_{2},i_{3}\right\} $ and $%
S=\left\{ s_{1},s_{2},s_{3}\right\} $, $q_{s}=1$ for all $s\in S,$ and 
\begin{gather*}
P_{i_{1}}:s_{2}\text{ }s_{1}\text{ }s_{3}\text{ }\emptyset ,\,P_{i_{2}}:s_{2}%
\text{ }s_{3}\text{ }s_{1}\text{ }\emptyset ,\,P_{i_{3}}:s_{3}\text{ }s_{2}%
\text{ }s_{1}\text{ }\emptyset ,\text{ } \\
\succ _{s_{1}}:\text{ }i_{1}\text{ }i_{3}\text{ }i_{2},\text{ }\succ
_{s_{2}}:i_{3}\text{ }i_{1}\text{ }i_{2},\text{ }\succ _{s_{3}}:i_{2}\text{ }%
i_{3}\text{ }i_{1}, \\
\succ _{s_{1}}^{\prime }:i_{1}\text{ }i_{2}\text{ }i_{3},\text{ }\succ
_{s_{2}}^{\prime }:i_{3}\text{ }i_{1}\text{ }i_{2},\text{ }\succ
_{s_{3}}^{\prime }:i_{2}\text{ }i_{1}\text{ }i_{3}, \\
\succ _{s_{1}}^{\prime \prime }:i_{1}\text{ }i_{3}\text{ }i_{2},\text{ }%
\succ _{s_{2}}^{\prime \prime }:i_{3}\text{ }i_{2}\text{ }i_{1},\text{ }%
\succ _{s_{3}}^{\prime \prime }:i_{2}\text{ }i_{1}\text{ }i_{3},
\end{gather*}%
Then, $\succ ^{\prime }$ and $\succ ^{\prime \prime }$ are improvements of $%
\succ $ for $\left\{ i_{1},i_{2}\right\} $ and, $\succ ^{\prime }$ is an
improvement of $\succ ^{\prime \prime }$ for $\left\{ i_{1},i_{2}\right\} $.
However, $\succ ^{\prime }$ does not more improve $\succ $ than $\succ
^{\prime \prime }$ for $\left\{ i_{1},i_{2}\right\} $, because $\left(
i_{1},i_{2}\right) \in \succ _{s_{2}}$ and $\left( i_{2},i_{1}\right) \in
\succ _{s_{2}}^{\prime \prime },$ but $\left( i_{1},i_{2}\right) \in \succ
_{s_{2}}^{\prime }$, which implies that (iii) does not hold. For $\mathcal{%
\vartriangleright }_{s}^{\prime }=\left\{ \succ _{s},\succ _{s}^{\prime
}\right\} $ and $\mathcal{\vartriangleright }_{s}^{\prime \prime }=\left\{
\succ _{s},\succ _{s}^{\prime \prime }\right\} $, 
\begin{eqnarray*}
m\left( \mathcal{\vartriangleright }_{s_{1}}^{\prime }\right) &:&i_{1}\text{ 
}\left[ i_{2}\text{ }i_{3}\right] ,\text{ }m\left( \mathcal{%
\vartriangleright }_{s_{1}}^{\prime \prime }\right) :i_{1}\text{ }i_{3}\text{
}i_{2}, \\
m\left( \mathcal{\vartriangleright }_{s_{2}}^{\prime }\right) &:&i_{3}\text{ 
}i_{1}\text{ }i_{2},\text{ }m\left( \mathcal{\vartriangleright }%
_{s_{2}}^{\prime \prime }\right) :i_{3}\text{ }\left[ i_{1}\text{ }i_{2}%
\right] , \\
m\left( \mathcal{\vartriangleright }_{s_{3}}^{\prime }\right) &=&m\left( 
\mathcal{\vartriangleright }_{s_{3}}^{\prime \prime }\right) :i_{2}\text{ }%
\left[ i_{1}\text{ }i_{3}\right] .
\end{eqnarray*}%
Then, the unique SOMSM for $\mathcal{\vartriangleright }^{\prime }$ is $\mu $
such that $\mu \left( i_{1}\right) =s_{1},$ $\mu \left( i_{2}\right) =s_{3}$
and $\mu \left( i_{3}\right) =s_{2}$. On the other hand, the unique SOMSM
for $\mathcal{\vartriangleright }^{\prime \prime }$ is $\mu ^{\prime }$ such
that $\mu ^{\prime }\left( i_{1}\right) =s_{1},$ $\mu ^{\prime }\left(
i_{2}\right) =s_{2}$ and $\mu ^{\prime }\left( i_{3}\right) =s_{3}$. Note
that $\mu ^{\prime }$ Pareto dominates $\mu $ for $\left\{
i_{1},i_{2}\right\} $, because $\mu \left( i_{1}\right) =\mu ^{\prime
}\left( i_{1}\right) =s_{1}$ and $\mu ^{\prime }\left( i_{2}\right)
=s_{2}P_{i_{2}}s_{3}=\mu \left( i_{2}\right) $.

Therefore, if $\phi $ is a student optimally M-stable mechanism, then $\phi
\left( \succ ,\succ ^{\prime }\right) =\mu $ and $\phi \left( \succ ,\succ
^{\prime \prime }\right) =\mu ^{\prime }$. However, since $\mu ^{\prime }$
Pareto dominates $\mu $ for $\left\{ i_{1},i_{2}\right\} $, $\phi $ must not
be strictly responsive to improvements. Therefore, contrary to the
responsiveness, the strict responsiveness is not compatible with the student
optimal M-stablility.

\subsection*{A.6. Proof of Theorem 2}

Let $\succ ^{\prime },\succ \in \mathcal{L}$ be such that $\succ ^{\prime }$
is an improvement of $\succ $ for $I^{\prime }$. For notational simplicity,
let 
\begin{equation*}
EA\left( M\left( \mathcal{\vartriangleright }^{\prime }\right) ,\succ
^{\prime }\right) =\mu ^{\ast }.
\end{equation*}

We show that $\mu ^{\ast }$ is an $I^{\prime }$\textbf{\ }optimal M-stable
matching for $\mathcal{\vartriangleright }^{\prime }.$ To show this, we
prove the following result.

\begin{claim}
We consider the EADA mechanism with $\left( M\left( \mathcal{%
\vartriangleright }^{\prime }\right) ,\succ ^{\prime }\right) $ (to derive $%
EA\left( M\left( \mathcal{\vartriangleright }^{\prime }\right) ,\succ
^{\prime }\right) )$. Fix any $k=1,\cdots ,K$. Suppose that $\nu $ is a
stable matching for $M\left( \mathcal{\vartriangleright }^{\prime }\right) $
that Pareto dominates $\hat{\mu}^{k}$ for $I^{\prime }$ and $\hat{\mu}%
^{k}\left( i^{\prime }\right) R_{i^{\prime }}\nu \left( i^{\prime }\right) $
for all $i^{\prime }\in \mathcal{E}^{k-1}$, where $\mathcal{E}^{0}=\emptyset 
$. Then, $\hat{\mu}^{k}\left( j\right) R_{j}\nu \left( j\right) $ for all $%
j\in \mathcal{E}^{k}(=\mathcal{E}^{k-1}\cup E^{k})$.
\end{claim}

\textbf{Proof.} Suppose not; that is, for some $k=1,\cdots ,K$, there is $%
i\in E^{k}$ such that $\nu \left( i\right) P_{i}\hat{\mu}^{k}\left( i\right) 
$; that is, $\kappa \left( i\right) =k$. Then, $\hat{\mu}^{k}\left( i\right) 
$ is underdemanded at $k$.

First, we show that there is a student $i_{d}\in I\setminus I^{\prime }$ and
a student $j_{d}$ such that $\hat{\mu}^{k}\left( i_{d}\right) P_{i_{d}}\nu
\left( i_{d}\right) $ and $\hat{\mu}^{k}\left( i_{d}\right) =\nu \left(
j_{d}\right) P_{j_{d}}\hat{\mu}^{k}\left( j_{d}\right) $. Let 
\begin{equation*}
S^{\ast }=\left\{ s\in S\text{ }\left\vert \text{ }s=\nu \left( i^{\prime
}\right) P_{i^{\prime }}\hat{\mu}^{k}\left( i^{\prime }\right) \text{ for
some }i^{\prime }\in I\right. \right\} \text{.}
\end{equation*}%
Then, by the assumption above, $\nu \left( i\right) \in S^{\ast }$. Since $%
\hat{\mu}^{k}\left( i^{\prime }\right) R_{i^{\prime }}\nu \left( i^{\prime
}\right) $ for all $i^{\prime }\in \mathcal{E}^{k-1}$ and $\hat{\mu}%
^{k}\left( i\right) $ is underdemanded at $k$, $\hat{\mu}^{k}\left( i\right)
\in S\setminus S^{\ast }$.

Let 
\begin{eqnarray*}
I_{1} &=&\left\{ i^{\prime }\in I\text{ }\left\vert \text{ }\nu \left(
i^{\prime }\right) \in S^{\ast }\text{ and }\hat{\mu}^{k}\left( i^{\prime
}\right) \in \left( S\cup \left\{ \emptyset \right\} \right) \setminus
S^{\ast }\right. \right\} , \\
I_{2} &=&\left\{ i^{\prime }\in I\text{ }\left\vert \text{ }\nu \left(
i^{\prime }\right) \in \left( S\cup \left\{ \emptyset \right\} \right)
\setminus S^{\ast }\text{ and }\hat{\mu}^{k}\left( i^{\prime }\right) \in
S^{\ast }\right. \right\} .
\end{eqnarray*}%
Then, $i\in I_{1}\setminus I_{2}$. We show $I_{2}\neq \emptyset $. Suppose
not; that is, $I_{2}=\emptyset $. Since $I_{1}\neq \emptyset $ and $%
I_{2}=\emptyset $, there is $s\in S^{\ast }$ such that $\left\vert \hat{\mu}%
^{k}\left( s\right) \right\vert <\left\vert \nu \left( s\right) \right\vert
\leq q_{s}$. Since $s\in S^{\ast }$, there is $i^{\prime }$ such that $s=\nu
\left( i^{\prime }\right) P_{i^{\prime }}\hat{\mu}^{k}\left( i^{\prime
}\right) $. Moreover, by Claim 3, $\hat{\mu}^{k}$ is stable for $M\left( 
\mathcal{\vartriangleright }^{\prime }\right) $. However, $\left\vert \hat{%
\mu}^{k}\left( s\right) \right\vert <q_{s}$ and $sP_{i^{\prime }}\hat{\mu}%
^{k}\left( i^{\prime }\right) $ contradict the nonwastefulness of $\hat{\mu}%
^{k}$. Hence $I_{2}\neq \emptyset $.

Then, we can let $i_{d}\in I_{2}$. Since $\nu \left( i_{d}\right) \notin
S^{\ast }$ and $\hat{\mu}^{k}\left( i_{d}\right) \in S^{\ast }$, $\hat{\mu}%
^{k}\left( i_{d}\right) P_{i_{d}}\nu \left( i_{d}\right) $. Let $\hat{\mu}%
^{k}\left( i_{d}\right) =s$. Since $s\in S^{\ast }$, there is $j_{d}$ such
that $s=\nu \left( j_{d}\right) P_{j_{d}}\hat{\mu}^{k}\left( j_{d}\right) $.
Since $\nu $ Pareto dominates $\hat{\mu}^{k}$ for $I^{\prime }$, $i_{d}\in
I\setminus I^{\prime }$.

Second, we show $\left( j_{d},i_{d}\right) \in \succ _{s}^{\prime }$. Since $%
\nu \left( j_{d}\right) =\hat{\mu}^{k}\left( i_{d}\right) =sP_{i_{d}}\nu
\left( i_{d}\right) $ and $\nu $ is stable for $M\left( \mathcal{%
\vartriangleright }^{\prime }\right) $, $\left( i_{d},j_{d}\right) \notin
m\left( \mathcal{\vartriangleright }_{s}^{\prime }\right) $; that is, both $%
\left( i_{d},j_{d}\right) \in \succ _{s}$ and $\left( i_{d},j_{d}\right) \in
\succ _{s}^{\prime }$ are not satisfied. Moreover, since $\succ _{s}^{\prime
}$ is an improvement of $\succ _{s}$ and $i_{d}\in I\setminus I^{\prime }$,
both $\left( j_{d},i_{d}\right) \in \succ _{s}$ and $\left(
i_{d},j_{d}\right) \in \succ _{s}^{\prime }$ are not satisfied. Since $\succ
_{s}$ is total, $\left( j_{d},i_{d}\right) \in \succ _{s}^{\prime }$.

Third, we show $\kappa \left( s\right) >k;$ that is, $s$ and $i_{d}$ have
not rejected at Round $k$ yet. Suppose not; that is, $\kappa \left( s\right)
\leq k$. By $\nu \left( j_{d}\right) =sP_{j_{d}}\hat{\mu}^{k}\left(
j_{d}\right) $ and Claim 1, $\nu \left( j_{d}\right) =sP_{j_{d}}\hat{\mu}%
^{\kappa \left( s\right) }\left( j_{d}\right) $. Since $s$ is underdemanded
at $\kappa \left( s\right) $, $\emptyset P_{j_{d}}^{\kappa \left( s\right)
}s $. Therefore, there is $j^{\prime }\in I$ such that $\kappa \left(
j^{\prime }\right) <\kappa \left( s\right) $, $sP_{j^{\prime }}\hat{\mu}%
^{k}\left( j^{\prime }\right) $ and $\left( j^{\prime },j_{d}\right) \in
m\left( \mathcal{\vartriangleright }_{s}^{\prime }\right) $. Since $\nu $ is
fair for $M\left( \mathcal{\vartriangleright }^{\prime }\right) ,$ $\nu
\left( j^{\prime }\right) R_{j^{\prime }}s=\nu \left( j_{d}\right) $ and
thus $\nu \left( j^{\prime }\right) P_{j^{\prime }}\hat{\mu}^{k}\left(
i^{\prime \prime }\right) $. However, since $j^{\prime }\in \mathcal{E}%
^{k-1} $, this contradicts $\hat{\mu}^{k}\left( i^{\prime }\right)
R_{i^{\prime }}\nu \left( i^{\prime }\right) $ for all $i^{\prime }\in 
\mathcal{E}^{k-1}$. Therefore, $s$ and $i_{d}$ have not eliminated at Round $%
k$ yet.

Moreover, since $s=\nu \left( j_{d}\right) P_{j_{d}}\hat{\mu}^{k}\left(
j_{d}\right) $, $j_{d}\notin \mathcal{E}^{k-1}$ and $\kappa \left(
j_{d}\right) \geq k$ is also satisfied. Therefore, in the SPDA of Round $k$, 
$i_{d}$ is accepted by $s$ but $j_{d}$ is not; that is, $j_{d}$ is rejected
by $s$ or $\emptyset P_{j_{d}}^{k}s$. Since $\left( j_{d},i_{d}\right) \in
\succ _{s}^{\prime }$, we have $\emptyset P_{j_{d}}^{k}s$ and hence there is 
$j^{\prime }$ such that $\kappa \left( j^{\prime }\right) <k$, $\left(
j^{\prime },j\right) \in m\left( \mathcal{\vartriangleright }_{s}^{\prime
}\right) $ and $sP_{j^{\prime }}\hat{\mu}^{k}\left( j^{\prime }\right) $.
Since $\nu $ is fair for $M\left( \mathcal{\vartriangleright }^{\prime
}\right) ,$ $\nu \left( j^{\prime }\right) R_{j^{\prime }}sP_{j^{\prime }}%
\hat{\mu}^{k}\left( j^{\prime }\right) $. However, since $j^{\prime }\in 
\mathcal{E}^{k-1}$, this contradicts $\hat{\mu}^{k}\left( i^{\prime }\right)
R_{i^{\prime }}\nu \left( i^{\prime }\right) $ for all $i^{\prime }\in 
\mathcal{E}^{k-1}$. \textbf{Q.E.D.}\newline

Now, we show Theorem 2. Suppose not; that is, $\nu $ Pareto dominates $\mu
^{\ast }$ for $I^{\prime }$. Then, by Claim 4, $\nu $ also Pareto dominates $%
\hat{\mu}^{k}$ for $I^{\prime }$ for all $k=1,\cdots ,K$. First, we have $%
\hat{\mu}^{1}\left( j\right) R_{j}\nu \left( j\right) $ for all $j\in 
\mathcal{E}^{1}\left( =E^{1}\right) $. Since $\hat{\mu}^{1}\left( j\right) =%
\hat{\mu}^{2}\left( j\right) $ for all $j\in \mathcal{E}^{1}$ and Claim 4 is
satisfied, $\hat{\mu}^{2}\left( j\right) R_{j}\nu \left( j\right) $ for all $%
j\in \mathcal{E}^{2}\left( =E^{2}\cup E^{1}\right) $. Likewise, we have $\mu
^{\ast }\left( j\right) =\hat{\mu}^{K}\left( j\right) R_{j}\nu \left(
j\right) $ for all $j\in \mathcal{E}^{K}=I$, but this contradicts that$\ \nu 
$ Pareto dominates $\mu ^{\ast }$ for $I^{\prime }$.

\subsection*{A.7. Proof of Theorem 3}

First, we show the following two results.

\begin{claim}
If all elements in $\mathcal{\vartriangleright }_{s}$ are linear orders,
then 
\begin{equation*}
m\left( \mathcal{\vartriangleright }_{s}\right) =\bigcap\limits_{\succ
_{s}\in \mathcal{\vartriangleright }_{s}}\succ _{s}\text{.}
\end{equation*}
\end{claim}

\textbf{Proof. }First, suppose 
\begin{equation*}
\left( i,j\right) \in m\left( \mathcal{\vartriangleright }_{s}\right) =\Pi
\left( \bigcup\limits_{\succ _{s}\in \mathcal{\vartriangleright }_{s}}\succ
_{s}\right) ,
\end{equation*}%
Then, $\left( j,i\right) \notin \bigcup\nolimits_{\succ _{s}\in \mathcal{%
\vartriangleright }_{s}}\succ _{s}$ and thus, $\left( j,i\right) \notin
\succ _{s}$ for all $\succ _{s}\in \mathcal{\vartriangleright }_{s}$. Since $%
\succ _{s}$ is total, $\left( i,j\right) \in \succ _{s}$ for all $\succ
_{s}\in \mathcal{\vartriangleright }_{s}$. Hence $\left( i,j\right) \in
\bigcap\nolimits_{\succ _{s}\in \mathcal{\vartriangleright }_{s}}\succ _{s}$.

Second, suppose%
\begin{equation*}
\left( i,j\right) \in \bigcap\limits_{\succ _{s}\in \mathcal{%
\vartriangleright }_{s}}\succ _{s},
\end{equation*}%
Then, there is no $\succ _{s}\in \mathcal{\vartriangleright }_{s}$ such that 
$\left( j,i\right) \in \succ _{s}$. Since $\succ _{s}$ is total, $\left(
i,j\right) \in \succ _{s}$ for all $\succ _{s}\in \mathcal{\vartriangleright 
}_{s}$. Hence $\left( i,j\right) \in m\left( \mathcal{\vartriangleright }%
_{s}\right) $. \textbf{Q.E.D.}\newline

\begin{claim}
If $\succ ^{\prime }$ more improves $\succ $ than $\succ ^{\prime \prime }$
for $I^{\prime }\subseteq I$, then for all $s\in S$,%
\begin{equation*}
m\left( \vartriangleright _{s}^{\prime }\right) \subseteq m\left(
\vartriangleright _{s}^{\prime \prime }\right) \text{.}
\end{equation*}
\end{claim}

\textbf{Proof. }By Claim 5, we have $m\left( \vartriangleright _{s}^{\prime
}\right) =\left( \succ _{s}\cap \succ _{s}^{\prime }\right) $ and $m\left(
\vartriangleright _{s}^{\prime \prime }\right) =\left( \succ _{s}\cap \succ
_{s}^{\prime \prime }\right) $. Let $i,j\in I$ be such that $\left(
i,j\right) \in \left( \succ _{s}\cap \succ _{s}^{\prime }\right) $. Then, $%
\left( i,j\right) \in \succ _{s}$ and $\left( i,j\right) \in \succ
_{s}^{\prime }$. We show $\left( i,j\right) \in \succ _{s}^{\prime \prime }$%
. Suppose not; that is, $\left( i,j\right) \notin \succ _{s}^{\prime \prime
} $. Since $\succ _{s}^{\prime \prime }\in \mathcal{L}$, $\left( j,i\right)
\in \succ _{s}^{\prime \prime }$. Since $\succ ^{\prime }$ is an improvement
of $\succ ^{\prime \prime }$ for $I^{\prime }$, $i\in I^{\prime }$. If $j\in
I\setminus I^{\prime }$, then $\left( i,j\right) \in \succ _{s}$ and $\left(
j,i\right) \in \succ _{s}^{\prime \prime }$ contradict $\succ ^{\prime
\prime }$ is an improvement of $\succ $ for $I^{\prime }$. Therefore, $j\in
I^{\prime }$. However, $\succ ^{\prime }$ more improves $\succ $ than $\succ
^{\prime \prime }$ for $I$, $\left( i,j\right) \in \succ _{s}$, $\left(
j,i\right) \in \succ _{s}^{\prime \prime }$, but $\left( i,j\right) \in
\succ _{s}^{\prime }$, which contradicts Condition (iii). Therefore, $\left(
i,j\right) \in \succ _{s}^{\prime \prime }$ and we have, for all $s\in S$, 
\begin{equation*}
m\left( \vartriangleright _{s}^{\prime }\right) =\left( \succ _{s}\cap \succ
_{s}^{\prime }\right) \subseteq \left( \succ _{s}\cap \succ _{s}^{\prime
\prime }\right) =m\left( \vartriangleright _{s}^{\prime \prime }\right) 
\text{.}
\end{equation*}%
\textbf{Q.E.D. }\newline

Now, we show Theorem 3. First, by Corollary 4, for any $\left( \succ ,\succ
^{\prime }\right) \in \mathcal{L}^{\left\vert S\right\vert }\times \mathcal{L%
}^{\left\vert S\right\vert }$, $\phi ^{\ast }\left( \succ ,\succ ^{\prime
}\right) $ is an SOMSM for $\vartriangleright ^{\prime }$.

Second, suppose that $\succ ^{\prime }$ is an improvement of $\succ $ for $%
I^{\prime }\subseteq I$. Then, by Theorem 2, $\phi ^{\ast }\left( \succ
,\succ ^{\prime }\right) $ is an $I^{\prime }$ optimally M-stable for $%
\vartriangleright ^{\prime }$. Therefore, $\phi ^{\ast }$ is improved-group
optimally M-stable.

Third, suppose that $\succ ^{\prime }$ more improves $\succ $ than $\succ
^{\prime \prime }$. Let $\mu $ be an M-stable matching for $%
\vartriangleright ^{\prime \prime }$. By Proposition 1 and Claim 5, $\mu $
is stable for $M\left( \vartriangleright ^{\prime \prime }\right) =\left(
\succ _{s}\cap \succ _{s}^{\prime \prime }\right) _{s\in S}$. By Claim 6, $%
\left( \succ _{s}\cap \succ _{s}^{\prime }\right) \subseteq \left( \succ
_{s}\cap \succ _{s}^{\prime \prime }\right) $ for all $s\in S$. Then, by
Remark 2, $\mu $ is also stable for $M\left( \vartriangleright ^{\prime
}\right) =\left( \succ _{s}\cap \succ _{s}^{\prime }\right) _{s\in S}$ and
thus M-stable for $\vartriangleright ^{\prime }$. Since $\phi ^{\ast }\left(
\succ ,\succ ^{\prime }\right) $ is student optimally M-stable for $%
\vartriangleright ^{\prime }$ and any M-stable matching for $%
\vartriangleright ^{\prime \prime }$ is also M-stable for $\vartriangleright
^{\prime }$, $\phi ^{\ast }\left( \succ ,\succ ^{\prime }\right) $ is not
Pareto dominated by any M-stable matching for $\vartriangleright ^{\prime
\prime }$. Hence $\phi ^{\ast }$ is responsive to improvements.

\subsection*{A.8. Failure of responsiveness in a wider domain}

In Section 6, we assume $\left( \succ ,\succ ^{\prime }\right) \in \mathcal{L%
}^{\left\vert S\right\vert }\times \mathcal{L}^{\left\vert S\right\vert }$
and show that $\phi ^{\ast }$ is responsive to improvements. Here, we show
that $\phi ^{\ast }$ is not responsive to improvements if $\left( \succ
,\succ ^{\prime }\right) \in \mathcal{W}^{\left\vert S\right\vert }\times 
\mathcal{W}^{\left\vert S\right\vert }$.

Suppose 
\begin{gather*}
P_{i_{1}}:s_{1}\text{ }s_{2}\text{ }s_{3}\text{ }\emptyset ,\,P_{i_{2}}:s_{1}%
\text{ }s_{2}\text{ }s_{3}\text{ }\emptyset ,\,P_{i_{3}}:s_{2}\text{ }s_{1}%
\text{ }s_{3}\text{ }\emptyset ,\text{ } \\
\succ _{s_{1}}:\text{ }i_{3}\text{ }i_{1}\text{ }i_{2},\text{ }\succ
_{s_{2}}:i_{1}\text{ }\left[ i_{2}\text{ }i_{3}\right] ,\text{ }\succ
_{s_{3}}:i_{3}\text{ }i_{1}\text{ }i_{2}, \\
\succ _{s_{1}}^{\prime }:i_{3}\text{ }i_{1}\text{ }i_{2},\text{ }\succ
_{s_{2}}^{\prime }:i_{1}\text{ }i_{2}\text{ }i_{3},\text{ }\succ
_{s_{3}}^{\prime }:i_{1}\text{ }i_{2}\text{ }i_{3}, \\
\succ _{s_{1}}^{\prime \prime }:i_{3}\text{ }i_{1}\text{ }i_{2},\text{ }%
\succ _{s_{2}}^{\prime \prime }:i_{1}\text{ }\left[ i_{2}\text{ }i_{3}\right]
,\text{ }\succ _{s_{3}}^{\prime \prime }:i_{1}\text{ }i_{3}\text{ }i_{2}.
\end{gather*}%
Then, $\succ _{s_{2}}$ and $\succ _{s_{2}}^{\prime \prime }$ are weak
orders. Thus, $\left( \succ ,\succ ^{\prime }\right) \in \mathcal{W}%
^{\left\vert S\right\vert }\times \mathcal{W}^{\left\vert S\right\vert }$
and $\succ ^{\prime }$ more improves $\succ $ than $\succ ^{\prime \prime }$
for $\left\{ i_{1},i_{2}\right\} \subseteq I$. Moreover, 
\begin{eqnarray*}
m\left( \mathcal{\vartriangleright }_{s_{1}}^{\prime }\right) &=&m\left( 
\mathcal{\vartriangleright }_{s_{1}}^{\prime \prime }\right) :i_{3}\text{ }%
i_{1}\text{ }i_{2} \\
m\left( \mathcal{\vartriangleright }_{s_{2}}^{\prime }\right) &:&i_{1}\text{ 
}i_{2}\text{ }i_{3},\text{ }m\left( \mathcal{\vartriangleright }%
_{s_{2}}^{\prime \prime }\right) :i_{1}\text{ }\left[ i_{2}\text{ }i_{3}%
\right] , \\
m\left( \mathcal{\vartriangleright }_{s_{3}}^{\prime }\right) &=&\left\{
\left( i_{1},i_{2}\right) \right\} ,\text{ }m\left( \mathcal{%
\vartriangleright }_{s_{2}}^{\prime \prime }\right) :\left[ i_{1}\text{ }%
i_{3}\right] \text{ }i_{2}\text{. }
\end{eqnarray*}%
Then, the unique SOMSM for $\mathcal{\vartriangleright }^{\prime }$ denoted
by $\mu ^{\prime }$ satisfies 
\begin{equation*}
\mu ^{\prime }\left( i_{1}\right) =s_{2},\text{ }\mu ^{\prime }\left(
i_{2}\right) =s_{3},\text{ and }\mu ^{\prime }\left( i_{3}\right) =s_{1}.
\end{equation*}%
On the other hand, an SOMSM for $\mathcal{\vartriangleright }^{\prime \prime
}$ denoted by $\mu ^{\prime \prime }$ satisfies 
\begin{equation*}
\mu ^{\prime \prime }\left( i_{1}\right) =s_{1},\text{ }\mu ^{\prime \prime
}\left( i_{2}\right) =s_{3},\text{ and }\mu ^{\prime \prime }\left(
i_{3}\right) =s_{2}.
\end{equation*}%
Since $\mu ^{\prime }$ is Pareto dominated for $\left\{ i_{1},i_{2}\right\} $
by $\mu ^{\prime \prime }$, $\phi ^{\ast }$ is not responsive to
improvements.

\subsection*{A.9. Double blocking pair}

Our stability notion seems similar to that of Fang and Yasuda (2021), but it
is different from ours. We briefly introduce their notion and explain the
difference between it and our notion.

Let $\succ ^{C}$ be the score priority order where $\left( i,j\right) \in
\succ ^{C}$ means that the score of $i$ is higher than that of $j$. Let $%
\succ _{s}^{P}$ be the preference priority order of $s$ where $\left(
i,j\right) \in \succ _{s}^{P}$ means that school $s$ prefers $i$ to $j$.
Fang and Yasuda (2021) define that a matching $\mu $ is said to \textbf{%
score }(resp.\textbf{\ preference})\textbf{\ blocked} by a student-school
pair $(i,s)$ if $\mu $ violates $\succ ^{C}$ (resp. $\succ _{s}^{P}$) of $%
i\notin \mu \left( s\right) $ for $s$. Moreover, for a matching $\mu $, a
student-school pair $(i,s)$ is said to be a \textbf{double blocking pair} if 
$\mu $ is both score and preference blocked by $(i,s)$.

Let $\vartriangleright _{s}=\left\{ \succ ^{C},\succ _{s}^{P}\right\} $ for
all $s\in S$. We show that an M-stable matching for $\vartriangleright $ may
have a double blocking pair. Suppose $I=\{i_{1},i_{2},i_{3}\}$ and $s\in S$,
where $\succ ^{C}:$ $i_{2}$ $i_{1}$ $i_{3}$, $\succ _{s}^{P}:$ $i_{3}$ $%
i_{1} $ $i_{2}$ and all students want to be matched to $s$. Let $\mu \left(
i_{2}\right) =\mu \left( i_{3}\right) =s$ and $\mu \left( i_{1}\right)
=\emptyset $. Then, $\mu $ is an M-stable matching for $\vartriangleright $,
because $i_{2}\succ ^{S}i_{1}$ and $i_{3}\succ _{s}^{P}i_{1}$. However, for $%
\mu $, $(i_{1},s)$ is a double blocking pair, because $\mu $ violates both $%
\succ ^{C}$ (over $i_{2}$) and $\succ _{s}^{P}$ (over $i_{3}$).

\end{document}